\documentclass[runningheads]{llncs}
\usepackage{graphicx}
\usepackage{underscore} 
\usepackage{amsmath}
\usepackage{amssymb}  
\usepackage{enumerate}
\usepackage{stmaryrd}
\usepackage{multirow}
\usepackage{color}
\usepackage{rotating}
\usepackage{verbatim}
\usepackage[colorlinks,
            linkcolor=blue,      
            anchorcolor=blue,  
            citecolor=blue,       
            ]{hyperref}

\newcommand{\invertedforall}{\begin{sideways}%
     \begin{sideways}$\forall$\end{sideways}\end{sideways}}
\newcommand{\invertedexists}{\begin{sideways}%
     \begin{sideways}$\exists$\end{sideways}\end{sideways}}

\newtheorem{fact}{Fact}

\newcommand{\set}[1]{\{#1\}}

\newcommand{\br}[1]{\llbracket #1\rrbracket}
  
\newcommand{\seq}[1]{\lbrack #1 \rbrack}
\newcommand{\dom}[1]{\mathsf{dom}( #1 )}

\renewcommand{\phi}{\varphi}
\newcommand{\sm}{\setminus}
\newcommand{\sub}{\subseteq}
\newcommand{\ve}{\varnothing}
\newcommand{\D}{\Diamond}
\newcommand{\B}{\Box}
\newcommand{\tup}[1]{\langle #1 \rangle}
\newcommand{\bd}{\blacklozenge}

\newcommand{\x}{\vec{x}}
\newcommand{\F}{\mathfrak{F}}
\renewcommand{\D}{\mathbb{D}}
\newcommand{\M}{\mathfrak{M}}
\renewcommand{\L}{\mathcal{L}}
\newcommand{\Lr}[1]{\llbracket{#1}\rrbracket}
\newcommand{\lr}[1]{\llp #1\rrp}
\newcommand{\md}{\models}
\newcommand{\rsto}{\upharpoonright}

\def\llp{\langle\mkern-9mu\langle\mkern2mu}
\def\rrp{\mkern2mu\rangle\mkern-9mu\rangle\mkern2.5mu}

\DeclareMathOperator{\sPa}{\mathsf{sPa}}
\DeclareMathOperator{\wPa}{\mathsf{wPa}}

\DeclareMathOperator{\Na}{\mathsf{Na}}

\newtheorem{notation}[definition]{Notation}
\begin{document}
\title{Reasoning about Dependence, Preference and Coalitional Power}
%
%\titlerunning{Abbreviated paper title}
% If the paper title is too long for the running head, you can set
% an abbreviated paper title here
%
\author{Qian Chen
\and
Chenwei Shi%\inst{1} 
\and
Yiyan Wang%\inst{1}
}
\authorrunning{Q.Chen, C. Shi \& Y. Wang}
% First names are abbreviated in the running head.
% If there are more than two authors, 'et al.' is used.
%
\institute{Tsinghua - Amsterdam Joint Research Centre for Logic, \\ Department of Philosophy, 
Tsinghua University, Beijing, China 
%\email{\quad scw@mail.tsinghua.edu.cn} \and
%Tsinghua - Amsterdam Joint Research Centre for Logic, Department of Philosophy, \\ Tsinghua University, Beijing, China \email{\quad wang-yy19@mails.tsinghua.edu.cn}
}
\maketitle              % typeset the header of the contribution
\begin{abstract}

This paper presents a logic of preference and functional dependence (LPFD) and its hybrid extension (HLPFD), 
both of whose sound and strongly complete axiomatization are provided. 
The decidability of LPFD are also proved.
The application of LPFD and HLPFD to modelling cooperative games in strategic and coalitional forms is explored. 
The resulted framework provides a unified view on Nash equilibrium, Pareto optimality and the core.
The philosophical relevance of these game-theoretical notions to discussions of collective agency is made explicit.
Some key connections with other logics are also revealed, for example, 
the coalition logic, the logic of functional dependence and the logic of ceteris paribus preference.

\end{abstract}

\section{Introduction}

Dependence, preference and coalitional power are three key concepts in game theory. 
There have been a lot of logical works on analyzing these three notions.  
To name but a few, for dependence, the dependence logic \cite{vaananen2007} has been studied in various ways (c.f.\cite{sep-logic-dependence})
and a simple logic of functional dependence is recently proposed in \cite{johan2021};
for coalitional power, the coalition logic \cite{pauly2002a} and the alternating-time temporal logic (ATL) \cite{alur2002ATL,goranko2004ATL} are representative; 
for preference, good surveys can be found in \cite{Hansson2002} and \cite[Chapter 1.1]{liu2011preference}. 
Despite not being explicitly emphasized, the concept of dependence permeates the analyses of the other two concepts, for example, in \cite{pauly2002a} and \cite{vanBenthem2007}.
However, as far as we know, there is hardly any logic explicitly modeling all of these three concepts, 
especially making dependence the hub to which the other two concepts join.
In this paper, we provide such a logic, which characterizes the interaction between the three concepts. 
Moreover, we show that by making the role of dependence explicit, 
our logical analysis leads to a unified view of several key concepts in game theory, namely Nash equilibrium, Pareto optimality and the core. 
We also explore a philosophical implication about collective agency of our logical analysis. 
We take the stability of a group to be an essential aspect of what makes it a coalition. 
Instead of focusing on intentionality as in the philosophical literature \cite{sep-shared-agency}, 
we elaborate on our understanding in a game theoretical context.

Our main work in this paper centers on introducing preference into the logic of functional dependence \cite{johan2021}
by adding preference relations in the original semantic model and a new modal operator in the original language for the intersection of different kinds of relations, 
including equivalence relations, preorders and strict preorders. 
By taking a game theoretic interpretation of the semantic setting, 
the new operator enables us to express not only Nash equilibrium but also Pareto optimality. 

While Nash equilibrium is taken to be a benchmark for modern logics of games and many logics have been demonstrated to be able to express it 
(see \cite[section 7.1]{vanBenthem2007} and the reference in it), 
Pareto optimality as an equally important notion in game theory
\footnote{For example, the prisoners' dilemma is the divergence between Nash equilibrium and Pareto optimality. } 
seems to receive less attention in logical literature than Nash equilibrium. 
As shown in this paper, to express Pareto optimality, the new modal operator is critical. 
In fact, given the operator, we can express a relativized version of Nash equilibrium and Pareto optimality,
that is, ``given the current strategies of some players, the current strategy profile of the other players would be a Nash equilibrium/Pareto optimality."
Moreover, by taking dependence relation into consideration, our logic shows that Nash equilibrium can be defined by Pareto optimality.

As Pareto optimality is seldom studied by logicians, 
compared to the \emph{non-cooperative} game theory, the \emph{cooperative} game theory \cite{peleg2007Introduction} seems not very salient to logicians either.
\footnote{The review on modal logic for games and information \cite[Chapter 20]{VANDERHOEK20071077} is exclusively about \emph{non-cooperative} game theory;
the book \cite{johanvanbenthem2014} touches on few issues on \emph{cooperative} game theory either.
The only exception we know is the work in \cite{agotnes2009}, 
where two different logics are proposed to reason about cooperative games. } 
We will demonstrate that our logic of preference and functional dependence (LPFD) can also be adapted to model 
a qualitative version of cooperative games in strategic and coalitional forms \cite[Section 11]{peleg2007Introduction}.
We will also show that a hybrid extension of LPFD can express the core, an essential solution concept in the cooperative games analogous to Nash equilibrium in the non-cooperative games. 
The core characterizes a coalition's stability as a state where none of its subcoalitions has any incentive to deviate even if they can.
The three concepts, dependence, preference and coalitional power,  crystallize in the core.
Through the lens framed by the three concepts, a unified view of the core, Nash equilibrium and Pareto optimality is revealed by our logics. 
%There are various ways of formulating the core. 
%Our formulation is akin to the one used in \cite{conzalez2021coreaxiom}, 
%which is in some sense more general than the one formulated in \cite{agotnes2009}.  
%More comparisons between our work and the logical work in \cite{agotnes2009} will come at the end of this paper.

In addition to the logics and their application to a unified analysis of key game theoretical concepts, 
our contributions include several technical results about the logics themselves.  
We provide a sound and strongly complete axiomatization respectively for LPFD and its hybrid extension (HLPFD). 
Moreover, we also prove that the satisfiability problem of LPFD is decidable. 
While the proof for the completeness result of HLPFD is standard,  
the completeness of LPFD is much harder to prove and requires new techniques. 
Our proof modifies the classical unraveling method \cite[Chapter 4.5]{modalLogic2001} and 
combines it with a special way of selecting the tree branches.

\subsubsection{The structure of the paper} is summarized as follows. 
The background on the logic of functional dependence (LFD) are presented in Section \ref{sec:background}. 
In the same section, we show how LFD can be used to analyze games in strategic form, 
especially the notion of coalitional effectiveness as modeled in \cite{pauly2002a}.
In Section \ref{sec:LPFD}, we introduce the logic of preference and functional dependence and 
show how it can naturally express Nash equilibrium and Pareto optimality. 
Section \ref{sec:axiomatization} contains sound and strongly complete axiomatization of LPFD and its hybrid extension 
and the decidability of LPFD's satisfiability problem.
For those who are not interested in the proof details, Section \ref{sec:completeLPFD} and Section \ref{sec:decidableLPFD} can be safely skipped. 
In Section \ref{sec:CGinLPFD}, we turn to our modelling of cooperative games in strategic and coalitional forms in LPFD and analyze the core. 
In Section \ref{sec:CAandPO}, we show how the core can be relevant to philosophical discussions of collective agency. 
Before conclusion, we compare our work with the logical works in \cite{agotnes2009} and \cite{vanBenthem2007}.

\subsubsection{Notations}

The following notations will be used throughout this paper. 
Let $A$ and $B$ be sets. Let $B^A$ denote the set of mappings from $A$ to $B$. 
Let $\mathcal{P}^{<\aleph_0}(A)$ denote the set of all finite subsets of $A$. 
We write $B\sub_{\aleph_0}A$ if $B\in\mathcal{P}^{<\aleph_0}(A)$. 
For each string $\x=(x_i:i\in I)$, we write $\mathsf{set}(\x)$ for the set $\set{x_i:i\in I}$.

\section{{LFD} for Coalitional Effectiveness}\label{sec:background}

In this section, we introduce LFD and make a first demonstration of its relevance to games.

\subsection{{LFD} Interpreted in Games}\label{sec:LFD}

LFD starts with a set of variables $\mathsf{V}$ and a domain of objects $O$. 
We take $\mathsf{V}$ as the set of players in a game and $O$ as the set of actions or strategies each player can take in the game. 
Then a set of admissible assignments of actions to players $A\subseteq O^\mathsf{V}$ can be collected to represent possible strategy profiles of the game.  
In addition, a relational vocabulary $(\mathsf{V},\mathsf{Pred},\mathsf{ar})$ is given to describe these possible strategy profiles, 
where $\mathsf{Pred}$ is a set of predicate symbols and $\mathsf{ar}: \mathsf{Pred}\rightarrow\mathbb{N}$ is an arity map, associating to each predicate $P\in \mathsf{Pred}$ a natural number $\mathsf{ar}(P)$.

In what follows, if there is no other explanation, the vocabulary $(\mathsf{V},\mathsf{Pred},\mathsf{ar})$ is the one such that $|\mathsf{V}|=\aleph_0$ and $|\set{P\in\mathsf{Pred}:\mathsf{ar}(P)=n}|=\aleph_0$ for each $n\in\omega$.

\begin{definition}[Dependence models]\label{def:Dmodel}
    %Given a vocabulary $(\mathsf{V},\mathsf{Pred},\mathsf{ar})$, a 
    A model is a pair $M=(O,I)$, where $O$ is a non-empty set of actions and $I$ is a mapping that assigns to each predicate $P\in \mathsf{Pred}$ a subset of $O^{\mathsf{ar}(P)}$. A dependence model $\mathbf{M}$ is a pair $\mathbf{M} = (M,A)$, where $M = (O,I)$ is a model and $A\subseteq O^\mathsf{V}$ is a set of strategy profiles. 

    For each $X\sub_{\aleph_0}\mathsf{V}$, we define a binary relation $=_X\sub A\times A$ such that $a=_Xa'$ if and only if $a{\rsto}X=a'{\rsto}X$, i.e., the action of $x$ in $a$ is the same as her action in $a'$ for each $x\in X$.
\end{definition} 
In a dependence model, when $A\neq O^\mathsf{V}$, some strategy profiles are missing. 
This gives rise to dependence between players' actions. 
Suppose a strategy profile $s$ for two players $x$ and $y$ is not in $A$.
Then $x$ and $y$ cannot act according to $s$ simultaneously. \footnote{
    In some sense this form of dependence is weak because it does not differentiate between different types of dependence, for example, correlation and causation. 
    However, the other side of the same coin is its generality which is helpful for capturing some common properties of different types of dependence. 
    For further explanation of how and what kinds of dependence can be captured in a dependence model, we refer readers to \cite{johan2021}. }
Differently, the standard setting of strategic form games usually contains all possible strategy profiles, namely $O^\mathsf{V}$.
This difference plays an essential role in making valid one of the axioms of the coalition logic, namely superadditivity, as we will explain in detail. 
For now, we turn to the syntax and semantics of LFD. To capture functional dependence, LFD uses two operators $\mathbb{D}$ and $D$ in its language. 
\begin{definition}
%Given a vocabulary $(\mathsf{V},\mathsf{Pred},\mathsf{ar})$, the language $\mathcal{L}$ is given by
The language $\mathcal{L}$ of \textnormal{LFD} is given by
$$\varphi :: = P\x \mid D_X y \mid \neg\varphi \mid \varphi\wedge\varphi \mid \mathbb{D}_X \varphi$$
where $P\in \mathsf{Pred}$, $\x = (x_1,\ldots, x_n)$ is a finite string of players of length $n = \mathsf{ar}(P)$, $X\subseteq_{\aleph_0} \mathsf{V}$ is a finite set of players and $y\in \mathsf{V}$ is a player. 
\end{definition}
$\mathbb{D}_X \varphi$ says that whenever the players in $X$ take their current actions, $\varphi$ is the case;  
$D_X y$ says that whenever the players in $X$ take their current actions, $y$ also takes its current action.
\begin{definition}\label{def:semantics-LFD}
Truth of a formula $\varphi\in \mathcal{L}$ in a dependence model $\mathbf{M} = (M,A)$ at a strategy profile $a\in A$ is defined as follows:
\begin{center}
\begin{tabular}{lll}
$\mathbf{M},a\models P\x$ & iff\ \ \  & $a(\x)\in I(P)$\\
$\mathbf{M},a\models D_X y$ & iff\ \  & $ a(y)=a'(y)$ for all $a'\in A$ with $a =_X a'$\\
$\mathbf{M},a\models \neg \varphi$ & iff\ \  & $\mathbf{M},a\not\models \varphi$\\
$\mathbf{M},a\models \varphi\wedge \psi$\ \  & iff\ \  & $\mathbf{M},a\models \varphi$ and $\mathbf{M},a\models \psi$\\
$\mathbf{M},a\models \mathbb{D}_X\varphi$ & iff\ \  & $\mathbf{M}, a'\models \varphi$ for all $a'\in A$ with $a =_X a'$ \\
\end{tabular}
\end{center}
%where $a =_X a'$ is the abbreviation for $a{\rsto}X=a'{\rsto}X$, which means  and $a=_y a'$ the abbreviation for $a=_{\set{y}} a'$.
\end{definition}
Note that $=_X$ is an equivalence relation on $A$ and $a=_\emptyset a'$ holds for all $a,a'\in A$.
So $\mathbb{D}_\emptyset$ is a universal operator and
we define $\invertedforall \varphi := \mathbb{D}_\emptyset \varphi$ and $\invertedexists \varphi := \neg \invertedforall \neg \varphi$.

\subsection{Effective Function and Coalition Logic in {LFD}}\label{sec:coalitionlogic}

We have introduced the basics of LFD and interpreted it in the setting of games in strategic form. 
In this subsection, we continue to explore the potential of this game theoretic perspective on LFD.
In particular, we show how the notion of coalitional effectiveness as modeled in \cite{pauly2002a} can be characterized in LFD.

The coalitional effectiveness that the coalition logic aims to reason about is formally characterized by an effectivity function $E_G$.
Based on this effectivity function, the main operator of the coalition logic $[C]\varphi$ is defined, 
expressing that the set of agents $C$ can force $\varphi$ to be the case at their current state. 

The effective function, when adapted in a dependence model $\mathbf{M} = (M,A)$, can be defined as
$E_\mathbf{M}: \mathcal{P}^{<\aleph_0}(\mathsf{V})\rightarrow \mathcal{P}(\mathcal{P}(A))$ satisfying
$$ S\in E_\mathbf{M}(X) \text{ iff } \exists a\in A, \forall a'\in A \text{ if } a' =_X a  \text{ then } a'\in S\enspace .$$
Here, $S\in E_\mathbf{M}(X)$ means that the coalition $X$ can force the game to be in $S$. 
We can express $S\in E_\mathbf{M}(X)$ in LFD as $\invertedexists \mathbb{D}_X\varphi$ assuming that $S = \br{\varphi}$, because
$$\mathbf{M}\models \invertedexists \mathbb{D}_X\varphi \text{ iff } \br{\varphi}\in E_\mathbf{M}(X)\enspace .$$
The operator $[C]\varphi$ in the coalition logic essentially has the same semantic meaning 
despite being interpreted in the neighborhood semantics.

We will not go into a detailed comparison between LFD and the coalition logic,
but only point out a substantial difference between $\invertedexists \mathbb{D}_X\varphi$ and $[C]\varphi$ 
with regard to the characteristic axiom of the coalition logic, superadditivity:
$$([C_1]\varphi_1\wedge [C_2]\varphi_2)\rightarrow [C_1\cup C_2](\varphi_1\wedge\varphi_2) \text{ where } C_1\cap C_2 = \emptyset\enspace .$$
Superadditivity fails for $\invertedexists \mathbb{D}_X\varphi$, 
because in a dependence model $A$ is not required to be $O^\mathsf{V}$ as we have noted after Definition \ref{def:Dmodel}.
In fact, the following proposition holds, which reveals that dependence between the players' actions invalidates superadditivity of coalitional effectiveness.  
\begin{proposition}
Let $\mathcal{M}$ be a class of dependence models. Superadditivity for $\invertedexists \mathbb{D}_X$, $(\invertedexists\mathbb{D}_X\varphi_1\wedge \invertedexists\mathbb{D}_Y\varphi_2)\rightarrow \invertedexists\mathbb{D}_{X\cup Y}(\varphi_1\wedge\varphi_2) \text{ where } X\cap Y = \emptyset$, is valid in $\mathcal{M}$ if and only if 
, $\set{a{\rsto}X:a\in A}=O^X$ for all $X\sub_{\aleph_0}\mathsf{V}$ and $((O,I),A)\in\mathcal{M}$.
%for any finite sets $X,Y\subseteq \mathsf{V}$ with $X\cap Y = \emptyset$ and any $a,a'\in A$, there is $a''\in A$ such that $a =_X a'' =_Y a'$.

% the class of dependence models whose available assignments $A$ satisfy the following property:
% for any finite sets $X,Y\subseteq \mathsf{V}$ with $X\cap Y = \emptyset$ and any $a,a'\in A$, there is $a''\in A$ such that $a =_X a'' =_Y a'$. 
% Reversely, if this property does not holds for a set of assignment $A$, then there must be a vocabulary $(pred,ar)$ and an interpretation map $I$ 
% such that superadditivity for $\invertedexists\mathbb{D}_X$ is not valid in $((O,I),A)$ 
\end{proposition}

As the readers who are familiar with the coalition logic can verify, except for superadditivity, its other axioms are all valid for $\invertedexists \mathbb{D}_X$
in LFD. In this sense, LFD provides a suitable framework for analyzing and understanding the relationship between dependence and coalitional effectiveness in games. 
However, as a framework for reasoning about other aspects of games, LFD and the coalition logic are both in want of a key element, namely the players' preference. 
In the next section, we extend LFD with the players' preference relations, study the resulted logic and 
show how it can capture key concepts in game theory. 
In Section \ref{sec:CGinLPFD}, the issue of coalitional power will come back 
and manifest itself in our analysis of cooperative games in strategic and coalitional forms.

\section{Logic of Preference and Functional Dependence}\label{sec:LPFD}

In this section, we extend LFD to LPFD. 

\subsection{Syntax and Semantic for LPFD}

\begin{definition}[Syntax]\label{def:PDlanguage}
    %Given a vocabulary $\text{Vo}=(\mathsf{V},\mathsf{Pred},ar)$, t
    The language $\L^\preceq$ of LPFD is given by:
    \[
        \L^\preceq\ni \phi ::= P\x \mid D_Xy \mid \neg\phi \mid \phi\wedge\phi \mid \Lr{X,Y,Z}\phi
    \]
    which only differs from the language of LFD in the new operator $\Lr{X,Y,Z}\phi$. In~$\L^\preceq$, $\mathbb{D}_X\varphi$ is defined as $\Lr{X,\emptyset,\emptyset}\phi$. We define $\lr{X,Y,Z}\phi:=\neg\Lr{X,Y,Z}\neg\phi$ and $D_XY:=\bigwedge_{y\in Y}D_Xy$ for each $Y\sub_{\aleph_0}\mathsf{V}$. 
\end{definition}
$\Lr{X,Y,Z}$ is an operator for ceteris paribus group preference, which is semantically interpreted as follows. 

\begin{definition}[PD-models]\label{def:PDmodel}%[Preference-dependence models]
    A {\em preference dependence model (PD-model)} is a pair $\mathbb{M}=(\mathbf{M},\preceq)$ in which $\mathbf{M}=(M,A)$ is a model and $\preceq:V\to\mathcal{P}(A\times A)$ is a mapping assigning to each $x\in \mathsf{V}$ a pre-order $\preceq_x$ on $A$.
    % and $s\precnsim_X t$ for ``$s\preceq_X t$ and $t\npreceq_X s$".\footnote{ 
    % Note that $\prec_x$ is the same as $\precnsim_{\set{x}}$, but $\prec_X$ is different from $\precnsim_X$.
    % }
\end{definition}
For each $x\in \mathsf{V}$, we define the binary relation $\prec_x=\{(a,b)\in\preceq_x:(b,a)\not\in\preceq_x\}$. For all $a,b\in A$, we write $a\preceq_Xb$($a\prec_Xb$) if $a\preceq_xb$($a\prec_xb$) for each $x\in X$. We write $s\simeq_X t$ if $s\preceq_X t$ and $t\preceq_X s$.

\begin{definition}
    Truth of PD-formulas of the form $P\x,D_Xy,\neg\phi$ or $\phi\wedge\psi$ is defined as in Definition \ref{def:semantics-LFD}.
    For formulas of the form $\Lr{X,Y,Z}\phi$, we say $\Lr{X,Y,Z}\phi$ is true at $a$ in $\mathbb{M}$, notation: $\mathbb{M},a \md\Lr{X,Y,Z}\phi$, if $\mathbb{M},a' \models \varphi$ for all $a'\in A$ satisfying $a =_{X} a'$, $a \preceq_{Y} a'$ and $a\prec_{Z} a'$.
    %  is defined as follows:
    % \begin{center}
    %     \begin{tabular}{lll}
    %     $\mathbb{M},a \md\Lr{X,Y,Z}\phi, \text{ if and only if, }$ \\ \qquad 
    %     $\mathbb{M},a' \models \varphi$ for all $a'\in A$ satisfying $a =_{X} a'$, $a \preceq_{Y} a'$ and $a\prec_{Z} a'$\\
    %     \end{tabular}
    % \end{center}

    A formula $\phi\in\L^\preceq$ is valid if $\mathbb{M},a\md\phi$ for all PD-model $\mathbb{M}=(M,A,\preceq)$ and $a\in A$. Let $\mathsf{LPFD}$ denote the set of all valid formulas in $\L^\preceq$.
\end{definition}

Note that $\br{\emptyset,\set{x},\emptyset} \varphi$ and $\br{\emptyset,\emptyset,\set{x}} \varphi$ are standard modal operators 
defined on $\preceq_x$ and $\prec_x$ respectively. 
Thus $\br{X,Y,Z} \varphi$ is in fact a standard modal operator 
defined on the intersection of the relations $=_X$, $\preceq_{Y}$ and $\prec_{Z}$. 

There are two types of interdependence between players in a game captured by LPFD. 
The first type, which comes from restricting what a player can do, is captured by the operators $\mathbb{D}_X$ and $D_X$; 
the second type, captured by $\br{X,Y,Z}$, concerns how the preferences of the players in $Y$ and $Z$ depend on the actions of the players in $X$. 
There is a close connection between LPFD and the work in \cite{vanBenthem2007} on ceteris paribus preference.
We will discuss this connection in Section 8. 
Next, we show how some key game theoretical notions can be expressed in LPFD.

\subsection{Pareto Optimality and Nash Equilibrium in LPFD}\label{sec: POandNEinLPFD}
Having laid out the basics of LPFD, we turn to questions concerning expressing and reasoning about Pareto optimality and Nash equilibrium in LPFD. 
One important assumption we will adopt is that the group of players $\mathsf{V}$ has to be finite. In LPFD, there is no such restriction on $\mathsf{V}$. 
However, it is worth noting that in the language of LPFD, all subscripts in the two operators need to be finite. 
So to express something like $\br{-X,\emptyset,X} \varphi$ in LPFD where $-X := V-X$, which is frequently referred to in game theory,  
we have to ensure that $X$ and $-X$ are both finite.

We start with recalling what Nash equilibrium and weak/strong Pareto optimality mean.
\begin{definition}\label{def:NwsP}
Let $\mathbb{M}$ be a PD-model and $X\sub \mathsf{V}$. Given that the players in $-X$ have acted according to the strategy profile $s\in A$,
\begin{itemize}
\item  $s$ is a \textbf{Nash equilibrium} for $X$ if for all $x\in X$ there is no $t=_{-\set{x}} s$ such that $s\prec_x t$;
\item  $s$ is \textbf{strongly Pareto optimal} for $X$ if there is no $t=_{-X} s$ such that 
(a) for all $x\in X$, $s\preceq_x t$ and (b) there is one $x\in X$ such that $s\prec_x t$;
\item  $s$ is \textbf{weakly Pareto optimal} for $X$  if there is no $t=_{-X} s$ such that for all $x\in X$, $s\prec_x t$.
\end{itemize}
\end{definition}
Note that such a way of defining the notions of Nash equilibrium, weak and strong Pareto optimality in a PD-model applies to all subgroups of $\mathsf{V}$ rather than only the whole group of players $\mathsf{V}$. 

It is relatively easy to get how Nash equilibrium and weak Pareto optimality can be expressed in LPFD, as the following fact shows.
\begin{fact}\label{fact:WPandNA}
Let $\mathbb{M}=(M,A,\preceq)$ be a PD-model and $s\in A$. Then
\begin{itemize}
\item $s$ is a Nash equilibrium for $X\subseteq \mathsf{V}$ given that the players in $-X$ have acted according to $s$, if and only if,
$\mathbb{M}, s\models \bigwedge_{x\in X}\br{-\set{x},\emptyset,\set{x}}\bot$;
\item $s$ is weakly Pareto optimal for $X\subseteq \mathsf{V}$ given that the players in $-X$ have acted according to $s$, if and only if,
$\mathbb{M}, s\models \br{-X,\emptyset,X}\bot$.
\end{itemize}
\end{fact}
In the case of weak Pareto optimality, 
because the truth condition of the operator  $\br{-X,\emptyset,X}$ depends on 
what formulas are satisfied on all elements in the set $\set{t\in A\mid s=_{-X} t, s\prec_X t}$, 
if it is an empty set and thus $\bot$ can be vacuously satisfied on all elements in it, 
then $s$ is weakly Pareto optimal for $X$. 

To express strong Pareto optimality in LPFD, we need to express the following model theoretical fact, namely,
the set $\set{t\in A\mid s=_{-X} t, s\preceq_X t \text{ and }t\npreceq_X s}$ $= \bigcup_{x\in X}\set{t\in A\mid s=_{-X} t, s\preceq_{X-\set{x}} t,s \prec_{x} t}$ is empty. 

Since $s\models \br{-X,X-\set{x}, \set{x}}\bot$ iff $\set{t\in A\mid s=_{-X} t, s\preceq_{X-\set{x}} t,s\prec_{x} t} = \emptyset$, 
we can define strong Pareto optimality as follows. 
\begin{fact}\label{fact:SP}
In a PD-model $\mathbb{M}$,
$s$ is strongly Pareto optimal for $X\subseteq \mathsf{V}$ given that the players in $-X$ have acted according to $s$ 
iff $\mathbb{M}, s\models \bigwedge_{x\in X}\br{-X,X-\set{x}, \set{x}}\bot$.
\end{fact}
\noindent To facilitate our discussion, we define weak and strong Pareto optimality and Nash equilibrium in LPFD as
\begin{align}
    \wPa X &:= \br{-X,\emptyset,X}\bot\\
    \sPa X &:= \bigwedge_{x\in X}\br{-X,X-\set{x}, \set{x}}\bot\\
    \Na X &:= \bigwedge_{x\in X} \br{-\set{x},\emptyset,\set{x}}\bot
\end{align}
An easy but important observation is that Nash equilibrium is a special case of Pareto optimality.
\begin{theorem}
    $\Na X = \bigwedge_{x\in X} \sPa \set{x}  = \bigwedge_{x\in X} \wPa \set{x}$.
\end{theorem}

\section{Calculus of LPFD and its Hybrid Extension}\label{sec:axiomatization}

In this section, a Kripke style semantics of LPFD shall be introduced.
It is proved to be equivalent to the standard semantics in Section \ref{sec:LFD}.
The new semantics provides us with a modal view, 
which facilitates our calculus $\mathsf{C}_\text{LPFD}$ and the proof of its soundness and strongly completeness.
We show that LPFD is decidable while it lacks the finite model property. 
Moreover, we extend it with nominals and give also a sound and complete calculus $\mathsf{C}_\text{HLPFD}$. 
In Section \ref{sec:CAandPO}, this hybrid extension will be useful in expressing a key game theoretic concept.

\subsection{Kripke Style Semantics}\label{sec:semEqu}

In this part, we introduce the Kripke style semantics for LPFD and 
show that it is equivalent to the standard semantics.

    \begin{definition}
        A {\em relational PD-frame (RPD-frame)} is a pair $\F=(W,\sim,\leq)$, where $W$ is a non-empty set, $\sim:V\to\mathcal{P}(W\times W)$ and $\leq:V\to\mathcal{P}(W\times W)$ are maps such that $\sim_x$ is an equivalence relation and $\leq_x$ is a pre-order for all $x\in \mathsf{V}$. For all $x\in \mathsf{V}$ and $X,Y,Z\sub_{\aleph_0}\mathsf{V}$, let $<_x=\{(w,u)\in\leq_x:(u,w)\not\in\leq_x\}$ and
        \begin{center}
            $R(X,Y,Z)=\bigcap_{x\in X}\sim_x\cap\bigcap_{y\in Y}\leq_y\cap\bigcap_{z\in Z}<_z.$
        \end{center}
        A {\em relational PD-model (RPD-model)} is a pair $\M=(\F,V)$ where $\F=(W,\sim,\leq)$ is a RPD-frame and $V$ is a valuation associating to each formula of the form $P\x$ a subset $V(P\x)$ of $W$. The valuation $V$ is required to satisfy the following condition for all $w,u\in W$ and $P\in\mathsf{Pred}$:
            \begin{center}
                if $w\sim_{\mathsf{set}(\x)}u$, then $w\in V(P\x)$ if and only if $u\in V(P\x)$.\ \ \ \ (Val)
            \end{center}
        Truth of a formula $\phi\in\L^\preceq$ in $\M=(W,\sim,\leq,V)$ at $w\in W$ is defined as follows:
        % \begin{align*}
        %     \M,w&\md P\x, \text{ iff } w\in V(P\x);\\
        %     \M,w&\md\neg\phi, \text{ iff } \M,w\not\md\phi;\\
        %     \M,w&\md\phi\wedge\psi, \text{ iff } \M,w\md\phi \text{ and } \M,w\md\psi;\\
        %     \M,w&\md D_Xy, \text{ iff } w\sim_yv \text{ for all } v\sim_Xw;\\
        %     \M,w&\md\Lr{X,Y,Z}\phi, \text{ iff } \M,v\md\phi \text{ for all } v\in R(X,Y,Z)(w).
        % \end{align*}
        \begin{center}
            \begin{tabular}{lll}
            $\M,w\models P\x$ & iff\ \ \  & $w\in V(P\x)$\\
            $\M,w\models D_X y$ & iff\ \  & $w\sim_yv \text{ for all } v\sim_Xw$\\
            $\M,w\models \neg \varphi$ & iff\ \  & $\M,w\not\models \varphi$\\
            $\M,w\models \varphi\wedge \psi$\ \  & iff\ \  & $\M,w\models \varphi$ and $\M,w\models \psi$\\
            $\M,w\models\Lr{X,Y,Z}\phi\ $ & iff\ \  & $\M,v\md\phi \text{ for all } v\in R(X,Y,Z)(w)$
            \end{tabular}
        \end{center}
        Validity is defined as usual. Let $\mathsf{RLPFD}$ denote the set of all valid $\L^\preceq$-formulas.
    \end{definition}

    \begin{proposition}\label{prop:m-rm}
        For each $\phi\in\L^\preceq$, if $\phi$ is satisfied by some PD-model $\mathbb{M}$, then it is satisfied by some RPD-model.
    \end{proposition}
    \begin{proof}
        Let $\mathbb{M}=(O,I,A,\preceq)$ be a PD-model. We define the RPD-model $rel(\mathbb{M})=(W,\sim,\leq,V)$ by $W:=A$, $V(P\x):=\{a\in A:a(\x)\in I(P)\}$ and $\sim_x:=(=_x)$, $\leq_x:=\preceq_x$ for each $x\in \mathsf{V}$. It is clear that for each $\phi\in\L^\preceq$ and $a\in A$, $\mathbb{M},a\md\phi$ if and only if $rel(\mathbb{M}),a\md\phi$. Then we are done.
    \end{proof}

    \begin{definition}
        Let $\M=(W,\sim,\leq,V)$ be a RPD-model. Then we define the PD-model $dp(\M)=(O,I,A,\preceq)$ induced by $\M$ as follows:
        \begin{itemize}
            \item $O=\{(x,|w|_x):x\in \mathsf{V},w\in W\text{ and } |w|_x=\{v\in W:w\sim_x v\}\}$.
            \item $A=\{w^*:w\in W\}$, where $w^*(x)=(x,|w|_x)$ for each $x\in \mathsf{V}$.
            \item $\preceq_x=\{(w^*,v^*):w\leq_xv\}$ for each $x\in \mathsf{V}$.
            \item $I$ is the interpretation maps each n-ary predicate $P$ to the set
                $$I(P)=\{w^*(\x):w\in W,\x\in \mathsf{V}^n\text{ and }w\in V(P\x)\}.$$
        \end{itemize}
        $I(P)$ is well-defined for each predicate $P$ since $x=y$ and $w\sim_xv$ whenever $w^*(x)=v^*(y)$. It is clearly that $\prec_x=\{(w^*,v^*):w<_xv\}$ for all $x\in \mathsf{V}$.
    \end{definition}

    \begin{proposition}\label{prop:rm-m}
        Let $\M$ be a RPD-model and $dp(\M)$ the PD-model induced by $\M$. Then for each $w$ in $\M$ and formula $\phi\in\L^\preceq$,
        \begin{center}
            $\M,w\md\phi$ if and only if $dp(\M),w^*\md\phi$.
        \end{center}
    \end{proposition}
    % \begin{proof}
    %     The proof proceeds by induction on the complexity of $\phi$. For the case $\phi=P\x$, suppose $\M,w\md\phi$. Then $w\in U(P\x)$ and $w^*(\x)\in I(P)$, which entails $dp(\M),w^*\md\phi$. Suppose $dp(\M),w^*\md\phi$. Then $w^*(\x)\in I(P)$, which entails $w\in U(P\x)$, i.e., $\M,w\md\phi$. The boolean case is trivial. The case when $\phi=D_Xy$ and $\phi=\Lr{X,Y,Z}\psi$ follow from the semantic definitions together with the fact that $w\sim_xv$ if and only if $w^*(x)=v^*(x)$ for each $x\in \mathsf{V}$.
    % \end{proof}
    By Proposition \ref{prop:m-rm} and Proposition \ref{prop:rm-m}, we have $\mathsf{RLPFD}=\mathsf{LPFD}$ immediately.
    % \begin{theorem}
    %     $\mathbf{RLPFD}=\mathbf{LPFD}$.
    % \end{theorem}

    \subsection{Hilbert-style Calculus ${\mathsf{C}_{\mathrm{LPFD}}}$}
    
    In this part, we present a calculus ${\mathsf{C}_{\mathrm{LPFD}}}$ of LPFD and show that ${\mathsf{C}_{\mathrm{LPFD}}}$ is sound, by which some key axioms are semantically explained. 

    \begin{enumerate}[(I)]
        \item[(Tau)] Axioms and rules for classical propositional logic;
        \item[(Nec)] from $\phi$ infer $\Lr{X,Y,Z}\phi$;
        \item[(K)] $\Lr{X,Y,Z}(\phi\to\psi)\to(\Lr{X,Y,Z}\phi\to\Lr{X,Y,Z}\psi)$;
        \item[] 
        \item[(Ord)] Axioms for preference relations:
            \item[(a)] $\Lr{X,Y,\ve}\phi\to\phi$;
            \item[(b)] $\lr{X,Y,Z}\lr{X',Y',Z'}\phi\to\lr{X\cap X', Y\cap Y',(Z\cap Y')\cup(Z\cap Z')\cup(Y\cap Z')}\phi$;
            \item[(c)] $\Lr{X,Y,Z}\phi\to\Lr{X',Y',Z'}\phi$, provided $X\sub X'$, $Y\sub Y'$ and $Z\sub Z'$.
            \item[(d)] $\lr{X,Y,Z}\phi\to\lr{X,Y\cup Z,Z}\phi$;
            \item[(e)] $(\phi\wedge\lr{X,Y,Z}\psi)\to\lr{X,Y,Z}(\psi\wedge\lr{X,Y,\ve}\phi)\vee\bigvee_{y\in Y}\lr{X,Y,Z\cup\{y\}}\psi$.
        \item[]        
        \item[(Dep)] Axioms and rules for dependence:
            \item[(a)] $D_XX$;
            \item[(b)] $\phi\to\D_X\phi$, provided $\phi\in\mathsf{Atom}(X)=\set{P\x:\mathsf{set}(\x)\sub X}\cup\set{D_{Y}z:Y\sub X}$;
            %\footnote{For each $X\sub_{\aleph_0}\mathsf{V}$, we define $\mathsf{Atom}(X)=\set{P\x:\mathsf{set}(\x)\sub X}\cup\set{D_{Y}z:Y\sub X}.$}
            %\item[(b2)] $D_Yz\to\D_X D_Yz$, provided that $Y\sub X$;
            \item[(c)] $D_XS\wedge D_ST\to D_XT$;
            \item[(d)] $D_XS\wedge\Lr{S,Y,Z}\phi\to\Lr{X,Y,Z}\phi$.
    \end{enumerate}
    In what follows, we write $\mathsf{C}$ for ${\mathsf{C}_{\mathrm{LPFD}}}$ if there is no danger of confusion.

    \begin{theorem}[Soundness]
        For each $\phi\in\L^\preceq$, $\vdash_\mathsf{C}\phi$ implies $\phi\in\mathsf{LPFD}$.
    \end{theorem}
    \begin{proof}
        We take (Ord,b) and (Ord,e) as two examples, showing their validity and giving some intuitions.
        Other axioms and rules can be easily checked to be valid. Let $\M=(W,\sim,\leq,V)$ be a RPD-model and $w\in W$ a point.

        For (Ord,b), it characterizes some kind of generalized transitivity. 
        Suppose $\M,w\md\lr{X,Y,Z}\lr{X',Y',Z'}\phi$. 
        Then there are points $u,v\in W$ such that $u\in R(X,Y,Z)(w)$, $v\in R(X',Y',Z')$ and $\M,v\md\phi$. 
        Let $T=(Z\cap Y')\cup(Z\cap Z')\cup(Y\cap Z')$. 
        It is obvious that $w\sim_{X\cap X'}v$ and $w\leq_{Y\cap Y'}$ hold. 
        It suffices to show that $w<_{T}v$. Suppose $x\in Z\cap Y'$. Then $w\leq_xu$, $u\not\leq_xw$ and $u\leq_xv$. 
        By the transitivity of $\leq_x$, we see $w\leq_xv$ and $v\not\leq_xw$, i.e., $w<_xv$. 
        Similarly, we see $w<_xv$ whenever $x\in Y\cap Z'$ or $x\in Z\cap Z'$. Hence $\M,w\md(Ord,b)$.

        For (Ord,e), it characterizes to some degree the definition of $<$. Suppose $\M,w\md\phi\wedge\lr{X,Y,Z}\psi$. Then there is a point $u\in R(X,Y,Z)(w)$ such that $\M,u\md\psi$. If $u\leq_Yw$, then clearly $\M,u\md\psi\wedge\lr{X,Y,\ve}\phi$, which entails $\M,w\md\lr{X,Y,Z}(\psi\wedge\lr{X,Y,\ve}\phi)$. Suppose $u\not\leq_Yw$. Then there is $y\in Y$ such that $u\not\leq_yw$ and so $w<_yu$. Recall that $u\in R(X,Y,Z)(w)$, we obtain $u\in R(X,Y,Z\cup\set{y})$ and so $\M,w\md\lr{X,Y,Z\cup\{y\}}\psi$. Hence $\M,w\md(Ord,e)$.
    \end{proof}

    \subsection{Strong Completeness of ${\mathsf{C}_{\mathrm{LPFD}}}$}\label{sec:completeLPFD}

    For the proof of completeness, a special kind of unraveling method is used. 
    The main reason we take such a method is that the `canonical model' need not be an RPD-model, and modification is needed. 
    To construct an RPD-model satisfying some given consistent set of formulas, we first pick out those so-called saturated formulas, 
    which are sufficient to determine the preference relations in the model. 
    Then we take `paths' as the domain of the desired model instead of using just maximal consistent sets, 
    which helps us deal with the intersections of relations. 
    The relations in this model are closures of some `one-step' relations, 
    which help solve the problems that arise from dependence formulas. 
    With such a model, we prove the Truth Lemma and so the Completeness Theorem.

    To define a model for some satisfiable set of formulas $\Gamma$, we first define the canonical quasi-frame and investigate some properties of it:

    \begin{definition}[Canonical Quasi PD-Frame]
        Let $\Delta$ be a set of $\L^\preceq$-formulas. We say that $\Delta$ is consistent if $\Delta\not\vdash\bot$. We say that $\Delta$ is a maximal consistent set (MCS) if $\Delta$ is consistent and every proper extension of $\Delta$ is not consistent. The canonical Quasi PD-frame $\F^q=(W^q,R^q)$ of $\mathsf{C}$ is defined as follows:
        \begin{itemize}
            \item $W^q$ is the set of all MCSs;
            \item for all $X,Y,Z\sub_{\aleph_0} \mathsf{V}$, we define $R^q(X,Y,Z)\sub W^q\times W^q$ by:
            \begin{center}
                $wR^q(X,Y,Z)u$ if and only if $\{\phi\in\L^\preceq:\Lr{X,Y,Z}\phi\in w\}\sub u$.
            \end{center}            
        \end{itemize}
    \end{definition}

    \begin{proposition}\label{prop:Rq}
        For all $\Delta_1,\Delta_2,\Delta_3\in W^q$ and $X,Y,Z\sub_{\aleph_0}\mathsf{V}$:
        \begin{enumerate}[(1)]
            \item $R^q(X,Y,\ve)$ is reflexive;
            \item If $\Delta_1 R^q(X,Y,Z)\Delta_2$, then $\Delta_1 R^q(X',Y',Z')\Delta_2$ for all $X'\sub X$, $Y'\sub Y\cup Z$ and $Z'\sub Z$;
            \item For all $Z'\sub_{\aleph_0}\mathsf{V}$, if $Z,Z'\sub Y$, $\Delta_1 R^q(X,Y,Z)\Delta_2$ and $\Delta_2 R^q(X,Y,Z')\Delta_3$, then $\Delta_1 R^q(X,Y,Z\cup Z')\Delta_3$;
            \item If $D_XS\in\Delta_1$ and $\Delta_1 R^q(X,Y,Z)\Delta_2$, then $\Delta_1 R^q(S,Y,Z)\Delta_2$ and $D_XS\in\Delta_2$.
        \end{enumerate}
    \end{proposition}
    \begin{proof}
        (1) follows form axiom (Ord,a), (2) follows from Axiom (Ord,c,d), (3) follows from axiom (Ord,b) and (4) follows from axiom (Dep,b,d) immediately.
    \end{proof}

    \begin{definition}
        Let $\Sigma$ be a MCS and $\lr{X,Y,Z}\phi\in\Sigma$. We say that $\lr{X,Y,Z}\phi$ is a saturated formula in $\Sigma$ if $\bigvee_{y\in Y}\lr{X,Y,Z\cup\{y\}}\phi\not\in\Sigma$ and $Y\cap Z=\ve$. Let $S(\Sigma)$ denote the set of all saturated formulas in $\Sigma$.
    \end{definition}

    \begin{lemma}\label{lem:goodcut}
        Let $\Sigma\in W^q$ be a MCS, $\lr{X,Y,Z}\phi\in\Sigma$ and $S=Y\cup Z$. Then there is $T\in\mathcal{P}(\mathsf{V})$ such that $\lr{X,T,(Y\cup Z)\sm T}\phi\in S(\Sigma)$.
    \end{lemma}
    \begin{proof}
        The proof proceeds by induction on the size $n$ of $Y\sm Z$. When $n=0$, one obtains $Z=Y\cup Z$. By axiom (Ord,c), $\lr{X,\ve,Z}\phi\in\Sigma$. Note that $\bigvee\ve=\bot\not\in\Sigma$, $\ve$ is the desired set. Suppose $n>0$ and $Y\sm Z=\{y_0,\cdots,y_{n-1}\}$. If $\lr{X,Y,Z\cup\{y_i\}}\phi\not\in\Sigma$ for any $i<n$, then $T=Y\sm Z$ satisfies the requirement. Suppose $\lr{X,Y,Z\cup\{y_i\}}\phi\in\Sigma$ for some $i<n$. Then we see $|Y\sm(Z\cup\{y_i\})|<n$ and by induction hypothesis, there is $T\in\mathcal{P}(\mathsf{V})$ such that $\lr{X,T,(Y\cup Z)\sm T}\phi\in S(\Sigma)$. Since $Y\cup Z=Y\cup(Z\cup\{y_i\})$, $T$ satisfies the requirement.
    \end{proof}

    \begin{lemma}\label{lemma:repair}
        Let $\Sigma\in W^q$ and $\lr{X,Y,Z}\phi\in S(\Sigma)$. Then there is a MCS $\Delta\in R^q(X,Y,Z)(\Sigma)$ such that $\phi\in\Delta$ and $\Delta R^q(X,Y,\ve)\Sigma$.
    \end{lemma}
    \begin{proof}
        We write $\B$ for $\Lr{X,Y,Z}$ and $\bd$ for $\lr{X,Y,\ve}$ in this proof. It is sufficient to show that $\Delta_0=\{\psi:\B\psi\in\Sigma\}\cup\{\bd\gamma:\gamma\in\Sigma\}\cup\{\phi\}$ is consistent. 
        Otherwise, there are formulas $\B\psi_1,\cdots,\B\psi_n,\gamma_1,\cdots,\gamma_m,\in\Sigma$ such that
        $$\vdash\psi_1\wedge\cdots\wedge\psi_n\wedge\bd\gamma_1\wedge\cdots\wedge\bd\gamma_m\wedge\phi\to\bot.$$
        Let $\gamma=\gamma_1\wedge\cdots\wedge\gamma_m$ and $\psi=\psi_1\wedge\cdots\wedge\psi_n$. Clearly, $\gamma\in\Sigma$. By axiom (Nec) and (K), we have $\vdash\bd\gamma\to(\bd\gamma_1\wedge\cdots\wedge\bd\gamma_m$). Thus $\vdash\psi\wedge\bd\gamma\wedge\phi\to\bot$, which entails $\vdash\B\psi\to\neg\Diamond(\phi\wedge\bd\gamma)$. Note that $\B\psi\in\Sigma$, we have $\neg\Diamond(\phi\wedge\bd\gamma)\in\Sigma$. Since $\gamma\wedge\Diamond\phi\in\Sigma$ and $(\gamma\wedge\Diamond\phi)\to\Diamond(\phi\wedge\bd\gamma)\vee\bigvee_{y\in Y}\lr{X,Y,Z\cup\{y\}}\phi$ is an instant of axiom (Ord,e), we obtain $\bigvee_{y\in{Y}}\lr{X,{Y},{Z}\cup\{y\}}\in\Sigma$, which contradicts that $\lr{X,Y,Z}\phi\in S(\Sigma)$.
    \end{proof}

    With the help of Lemma \ref{lem:goodcut} and Lemma \ref{lemma:repair}, we are now able to define the paths in $W^q$, which constitute the domain of our desired model.

    \begin{definition}
        A path in $W^q$ is a sequence $\pi=\tup{\Sigma_0,\psi_0,\cdots,\Sigma_{n-1},\psi_{n-1},\Sigma_n}$ in which the following conditions hold for all $i<n$:
        \begin{itemize}
            \item $\psi_i=\lr{X_i,Y_i,Z_i}\phi_i\in S(\Sigma_i)$ is a saturated formula in $\Sigma_i\in W^q$;% for all $i<n$;
            \item $\phi_i\in\Sigma_{i+1}\in W^q$, $\Sigma_{i+1}R^q(X_i,Y_i,\ve)\Sigma_i$ and $\Sigma_i R^q(X_i,Y_i,Z_i)\Sigma_{i+1}$.
        \end{itemize}
        We denote $\Sigma_0$ by $\mathrm{start}(\pi)$, $\Sigma_n$ by $\mathrm{last}(\pi)$ and the set of all paths by $\mathrm{Path}$.% in $W^q$.
    \end{definition}

    In what follows, let $\Gamma$ be some fixed consistent set. Without loss of generality, suppose $\Gamma$ is a MCS. We now construct a model for $\Gamma$.

    \begin{definition}[$\Gamma$-Canonical PD-model]
        The $\Gamma$-canonical PD-model $\M^c_\Gamma=(\F^c_\Gamma,V^c)$, in which $\F^c_\Gamma=(W^c_\Gamma,\leq^c,\sim^c)$, is defined as follows:
        \begin{itemize}
            \item $W^c_\Gamma=\set{\pi\in\mathrm{Path}:\mathrm{start}(\pi)=\Gamma}$, and we write $W^c$ for $W^c_\Gamma$ in what follows;
            \item for all $y\in \mathsf{V}$ and $\pi,\pi'\in W^c$, $\pi\leq_y\pi'$ iff one of the following holds:
            \begin{itemize}
                \item $\pi'=\tup{\pi,\lr{X,Y,Z}\phi,\Sigma}$ and $y\in Y\cup Z$;
                \item $\pi=\tup{\pi',\lr{X,Y,Z}\phi,\Sigma}$ and $y\in Y$;
                \item $\pi=\pi'$.
            \end{itemize}
            Let $\leq^c_y$ be the transitive closure of $\leq_y$.

            \item for all $s\in \mathsf{V}$ and $\pi,\pi'\in W^c$, $\pi\rightharpoonup_s\pi'$ if and only if $\pi'=\tup{\pi,\lr{X,Y,Z}\phi,\Sigma}$ and $D_Xs\in\mathrm{last}(\pi)$. Let $\rightleftharpoons_s$ be the reflexive-symmetric closure of $\rightharpoonup_s$.
            
            Let $\sim^c_s$ be the transitive closure of $\rightleftharpoons_s$.

            \item for all $P\x\in\L$, $V^c(P\x)=\{\pi\in W^c: P\x\in\mathrm{last}(\pi)\}$.
        \end{itemize}
        For all $X,Y,Z\sub_{\aleph_0}\mathsf{V}$, the binary relations $R^c(X,Y,Z)$, $\sim^c_X$, $\leq^c_Y$ and $<^c_Z$ are defined in the natural way. By Axiom (Dep,a), $D_XX$ always holds. Thus for each $\pi\in W^c_\Gamma$ and $\pi'=\tup{\pi,\lr{X,Y,Z}\phi,\Delta}$, we have $\pi R^c(X,Y,Z)\pi'$.
    \end{definition}
    
    To characterize the structure of $W^c$, we define $T\sub W^c\times W^c$ as follows:
        \begin{center}
            $\pi T\pi'$ if and only if $\pi'$ is of the form $\tup{\pi,\lr{X,Y,Z}\phi,\Sigma}$.
        \end{center}
    It is clear that $(W^c,T)$ is a tree. Then for all $\pi,\pi'\in W^c$, there is a shortest $T$-sequence $\tup{\pi_0,\cdots,\pi_n}$ such that $\pi=\pi_0$, $\pi'=\pi_n$ and for all $i<n$, $\pi_i T\pi_{i+1}$ or $\pi_{i+1}T\pi_i$. We denote the shortest sequence by $T^\pi_{\pi'}$.

    \begin{fact}\label{fact:shortestroad}
        Let $\pi,\pi'\in W^c$, $T^\pi_{\pi'}=\tup{\pi_0,\cdots,\pi_n}$ and $y,s\in \mathsf{V}$. Then
        \begin{enumerate}[(1)]
            \item $\pi\sim^c_s\pi'$ iff $\pi_i\rightleftharpoons_s\pi_{i+1}$ for all $i<n$.
            \item $\pi\leq^c_y\pi'$ iff $\pi_i\leq_y\pi_{i+1}$ for all $i<n$.
        \end{enumerate}
    \end{fact}
    \begin{proof}
        Since $\rightleftharpoons_s,\leq_y\sub(T\cup T^{-1})$, $\sim^c_s$ is the transitive closure of $\rightleftharpoons_s$ and $\leq^c_y$ the transitive closure of $\leq_y$, the proof can be done by induction on $n$ easily.
    \end{proof}

    In what follows, we show that the relations $R^c(X,Y,Z)$ are consistent with the relations $R^q(X,Y,Z)$. 

    \begin{lemma}\label{lem:oneRcRq}
        Let $\pi,\pi'\in W^c$, $X,Y,Z\sub \mathsf{V}$, $\pi\rightleftharpoons_X\pi'$, $\pi\leq_Y\pi'$ and $\pi<_Z\pi'$. Then 
        \begin{enumerate}[(1)]
            \item $\mathrm{last}(\pi) R^q(X,Y,Z)\mathrm{last}(\pi')$.
            \item if $D_XS\in\mathrm{last}(\pi)$, then $\pi\rightleftharpoons_S\pi'$.
        \end{enumerate}        
    \end{lemma}
    \begin{proof}
        Suppose $\pi\rightleftharpoons_X\pi'$, $\pi\leq_Y\pi'$ and $\pi<_Z\pi'$. Then we have three cases:
        \begin{itemize}
            \item $\pi=\pi'$. Then $Z=\ve$. By Proposition \ref{prop:Rq}(1), $R^q(X,Y,\ve)$ is reflexive and $\mathrm{last}(\pi) R^q(X,Y,Z)\mathrm{last}(\pi')$.
            \item $\pi=\tup{\pi',\lr{X',Y',Z'}\psi,\Delta}$. Then $Z=\ve$, $Y\sub Y'$ and $D_{X'}X\in\mathrm{last}(\pi')$. Clearly, $\mathrm{last}(\pi')R^q(X',Y',Z')\mathrm{last}(\pi)$. by Proposition \ref{prop:Rq}(4), $D_{X'}X\in\mathrm{last}(\pi)$. Recall that one has $\mathrm{last}(\pi)R^q(X',Y',\ve)\mathrm{last}(\pi')$, by Proposition \ref{prop:Rq}(2,4), we see $\mathrm{last}(\pi) R^q(X,Y,Z)\mathrm{last}(\pi')$.
            \item $\pi'=\tup{\pi,\lr{X',Y',Z'}\psi,\Delta}$. Then $Z\sub Z'$, $Y\sub Y'\cup Z'$ and $D_{X'}X\in\mathrm{last}(\pi)$. Note that $\mathrm{last}(\pi)R^q(X',Y',Z') \mathrm{last}(\pi')$, by Proposition \ref{prop:Rq}(2,4), we see $\mathrm{last}(\pi') R^q(X,Y,Z)\mathrm{last}(\pi')$.
        \end{itemize}
        Hence $\mathrm{last}(\pi) R^q(X,Y,Z)\mathrm{last}(\pi')$ and (1) holds.

        For (2), suppose $D_XS\in\mathrm{last}(\pi)$. Then we have also three cases:
        \begin{itemize}
            \item $\pi=\pi'$. Note that $\rightleftharpoons_S$ is reflexive, $\pi\rightleftharpoons_S\pi'$.
            \item $\pi\rightharpoonup_X\pi'$. Then $\pi'$ is of the form $\tup{\pi,\lr{X',Y',Z'}\psi,\Delta}$ and $D_{X'}X\in\mathrm{last}(\pi)$. By axiom (Dep,c), $D_{X'}S\in\mathrm{last}(\pi)$. Thus $\pi\rightharpoonup_S\pi'$.
            \item $\pi'\rightharpoonup_X\pi$. Then $\pi$ is of the form $\tup{\pi',\lr{X',Y',Z'}\psi,\Delta}$ and $D_{X'}X\in\mathrm{last}(\pi')$. By (1), $D_XS\in\mathrm{last}(\pi')$. By axiom (Dep,c), $D_{X'}S\in\mathrm{last}(\pi)$. Thus $\pi'\rightharpoonup_S\pi$.
        \end{itemize}
        Hence $\pi\rightleftharpoons_S\pi'$ and (2) holds.
    \end{proof}

    \begin{lemma}\label{lem:RcRq}
        Let $\pi,\pi'\in W^c$, $X,Y,Z\sub_{\aleph_0} \mathsf{V}$ and $\pi R^c(X,Y,Z)\pi'$. Then 
        \begin{enumerate}[(1)]
            \item $\mathrm{last}(\pi) R^q(X,Y,Z)\mathrm{last}(\pi')$.
            \item $D_XS\in\mathrm{last}(\pi)$ implies $\pi R^c(S,Y,Z)\pi'$.
        \end{enumerate}
    \end{lemma}
    \begin{proof}
        Suppose $\pi R^c(X,Y,Z)\pi'$. Then $\pi\sim^c_X\pi'$, $\pi\leq^c_{Y\cup Z}\pi'$ and $\pi<^c_Z\pi'$. Let $T^\pi_{\pi'}=\tup{\pi_0,\cdots,\pi_n}$. By Fact \ref{fact:shortestroad}, for all $i<n$, $\pi_i\rightleftharpoons_X\pi_{i+1}$ and $\pi_i\leq_{Y\cup Z}\pi_{i+1}$. Moreover, for each $z\in Z$, there is $i_z\in n$ such that $\pi_{i_z}<_{z}\pi_{i_z+1}$. Then by Lemma \ref{lem:oneRcRq}(1), $\mathrm{last}(\pi_i) R^q(X,Y\cup Z,\ve)\mathrm{last}(\pi_{i+1})$ for all $i\in n$ and for all $z\in Z$, $\mathrm{last}(\pi_{i_z}) R^q(X,Y\cup Z,\{z\})\mathrm{last}(\pi_{i_z+1})$. Then by Proposition \ref{prop:Rq}(2,3), we see $\mathrm{last}(\pi) R^q(X,Y,Z)\mathrm{last}(\pi')$ and (1) holds. Suppose $D_XS\in\mathrm{last}(\pi)$. Note that $\pi\sim^c_X\pi_i$ for all $i\leq n$, by (1), $D_XS\in\mathrm{last}(\pi_i)$ for all $i\leq n$. Then by Lemma \ref{lem:oneRcRq}(2), $\pi_i\rightleftharpoons_S\pi_{i+1}$ for all $i\in n$, which entails $\pi\sim^c_S\pi'$.
    \end{proof}

    The final step is to show that $\M^c$ is a PD-model in which $\Gamma$ is satisfiable.

    \begin{lemma}
        $\M^c$ is a PD-model.
    \end{lemma}
    \begin{proof}
        It suffices to show that $V^c$ satisfies (Val). Let $\pi,\pi'\in W^c$ be points such that $\pi\sim^c_X\pi'$. By Lemma \ref{lem:RcRq},  $\mathrm{last}(\pi) R^q(X,\ve,\ve)\mathrm{last}(\pi')$. Assume $P\x\in\mathrm{last}(\pi)$, then by axiom (Dep,b), $\D_XP\x\in\mathrm{last}(\pi)$, which entails $P\x\in\mathrm{last}(\pi')$. Similarly, we can verify that $P\x\in\mathrm{last}(\pi')$ implies $P\x\in\mathrm{last}(\pi)$. Thus $V^c$ satisfies (Val) and so $\M^c_\Gamma$ is a PD-model.
    \end{proof}

    \begin{lemma}[Truth Lemma]
        For each formula $\phi\in\L^\preceq$ and path $\pi\in W^c$, $\M^c,\pi\md\phi$ if and only if $\phi\in\mathrm{last}(\pi)$.
    \end{lemma}
    \begin{proof}
        The proof proceeds by induction on the complexity of $\phi$. The case when $\phi$ is of the form $P\x$ is trivial. The Boolean cases are also trivial. Let $\phi$ be of the form $D_Xs$. Suppose $D_Xs\in\mathrm{last}(\pi)$. Let $\pi'\in W^c$ such that $\pi\sim^c_X\pi'$. By Lemma~\ref{lem:RcRq}, $\mathrm{last}(\pi) R^q(X,\ve,\ve)\mathrm{last}(\pi')$. Then by Proposition \ref{prop:Rq}(2,4), $\pi\sim^c_s\pi'$. Thus $\M^c,\pi\md D_Xs$. Suppose $D_Xs\not\in\mathrm{last}(\pi)$. Let $\pi'=\tup{\pi,\lr{X,\ve,\ve}\top,\mathrm{last}(\pi)}$. Then $\pi\not\rightharpoonup_s\pi'$ and so $\pi\not\rightleftharpoons_s\pi'$. Clearly, $T^\pi_{\pi'}=\tup{\pi,\pi'}$. By Fact \ref{fact:shortestroad}, $\pi\not\sim^c_s\pi'$. Note that $\pi\sim^c_X\pi'$, we see $\M^c,\pi\not\md D_Xs$. Let $\phi=\lr{X,Y,Z}\psi$. Suppose $\M^c,\pi\md\phi$. Then there is $\pi'\in R^c(X,Y,Z)\pi$ such that $\M^c,\pi'\md\psi$. By induction hypothesis, $\psi\in\mathrm{last}(\pi')$. By Lemma \ref{lem:RcRq}, $\mathrm{last}(\pi) R^q(X,Y,Z)\mathrm{last}(\pi')$. Then $\phi\in\mathrm{last}(\pi)$. Suppose $\phi\in\mathrm{last}(\pi)$. Without loss of generality, assume that $\phi\in S(\mathrm{last}(\pi))$. Then by Lemma~\ref{lemma:repair}, there is a $\Delta$ such that $\pi'=\tup{\pi,\phi,\Delta}$ is a path with $\psi\in\mathrm{last}(\pi')$. By induction hypothesis, $\M^c,\pi'\md\psi$. Note that $\pi R^c(X,Y,Z)\pi'$, we have $\M^c,\pi\md\phi$.
    \end{proof}

    \begin{theorem}
        For each $\Gamma\subseteq \L^\preceq$, if $\Gamma$ is consistent, then $\Gamma$ is satisfiable.
    \end{theorem}

    \subsection{Properties of LPFD}\label{sec:decidableLPFD}
    
    In this part, we prove that LPFD lacks the finite model property.
    The decidability of LPFD shall also be shown. 

    \begin{theorem}
        LPFD lacks the finite model property; that is, some formula $\phi\in\L^\preceq$ is only satisfiable in infinite RPD-models.
    \end{theorem}
    \begin{proof}
        Let $\phi=\neg(\Lr{\ve,\ve,\{z\}}\bot\vee\lr{\ve,\ve,\{z\}}\Lr{\ve,\ve,\{z\}}\bot)$. Note that for each PD-frame $\F=(W,\sim,\leq)$ and $z\in \mathsf{V}$, $<_z$ is irreflexive and transitive. Thus for each finite PD-frame $\mathfrak{G}$, we have  $\mathfrak{G}\md\Lr{\ve,\ve,\{z\}}\bot\vee\lr{\ve,\ve,\{z\}}\Lr{\ve,\ve,\{z\}}\bot$. Clearly, $\phi$ is satisfiable in $(\omega,\sim,\leq)$, where $\leq_z$ is the usual $\leq$ relation on $\omega$.
    \end{proof}

    In what follows, let $\alpha$ be some fixed formula, $\mathsf{V}_\alpha$ the set of variables occur in $\alpha$ and $\mathsf{Pred}_\alpha$ the set of predicates occur in $\alpha$. Without loss of generality, we assume that the modal depth of $\alpha$ is not 0. Then we define $\text{Vo}=(V_\alpha,\mathsf{Pred}_\alpha,ar{\rsto}V_\alpha)$ as the vocabulary restricted to $\alpha$. Let $\L_\alpha$ be the fragment of $\L^\preceq$ based on $\text{Vo}$, in which every formula is of modal degree no more than $\alpha$. It can be easily verified that up to modal equivalence, $\L_\alpha$ contains only finitely many formulas. 

    \begin{definition}
        A set $\Gamma$ of $\L_\alpha$-formulas is said to be a $\L_\alpha$-maximal consistent set if $\Gamma\not\vdash\bot$ and $\Gamma'\vdash\bot$ for all $\Gamma'$ such that $\Gamma\subsetneq\Gamma'\sub\L_\alpha$. Let $\mathrm{MCS}_\alpha$ denote the set of all $\L_\alpha$-maximal consistent sets. For all $X,Y,Z\sub V_\alpha$ and $\Delta,\Sigma\in\mathrm{MCS}_\alpha$, we write $\Delta R^p_{\alpha}(X,Y,Z)\Sigma$ if 
        \begin{center}
            $\{\lr{X\cap X',Y\cap Y',(Z\cap Y')\cup(Z'\cap Y)\cup(Z\cap Z')}\phi\in\L_\alpha:\lr{X',Y',Z'}\phi\in\Sigma\}\sub\Delta$.
        \end{center} 
    \end{definition}
    One may find that the definition of $R^p_{\alpha}(X,Y,Z)$ is modified from the Lemmon filtration. Given that $\lr{X,Y,Z}\phi\in\L_\alpha$ and $\Delta R^p_{\alpha}(X,Y,Z)\Sigma$, we see $\phi\in\Sigma$ implies $\lr{X,Y,\ve}\phi\in\Sigma$ and so $\lr{X,Y,Z}\phi\in\Delta$. Then we have the following proposition:
    
    \begin{proposition}\label{fact:MCSalpha}
        For all $\Delta_1,\Delta_2,\Delta_3\in\mathrm{MCS}_\alpha$ and $X,Y,Z\sub V_\alpha$,
        \begin{enumerate}[(1)]
            \item $R^p_{\alpha}(X,Y,\ve)$ is reflexive;
            \item If $\Delta_1 R^p_{\alpha}(X,Y,Z)\Delta_2$, then $\Delta_1 R^p_{\alpha}(X',Y',Z')\Delta_2$ for all $X'\sub X$, $Y'\sub Y\cup Z$ and $Z'\sub Z$;
            \item If $D_XS\in\Delta_1$ and $\Delta_1 R^p_{\alpha}(X,Y,Z)\Delta_2$, then $\Delta_1 R^p_{\alpha}(S,Y,Z)\Delta_2$ and $D_XS\in\Delta_2$
            \item For all $Z'\sub_{\aleph_0}V_\alpha$, if $Z,Z'\sub Y$, $\Delta_1 R^p_{\alpha}(X,Y,Z)\Delta_2$ and $\Delta_2 R^p_{\alpha}(X,Y,Z')\Delta_3$, then $\Delta_1 R^p_{\alpha}(X,Y,Z\cup Z')\Delta_3$.
        \end{enumerate}
    \end{proposition}
    \begin{proof}
        (1) and (2) are trivial. For (3), $\Delta_1 R^p_{\alpha}(S,Y,Z)\Delta_2$ follows from axiom (Dep,d). Recall that the modal depth of $\alpha$ is not 0, we see $\D_XD_XS\in\Delta_1$ and so $D_XS\in\Delta_2$. For (4), suppose $\lr{X_0,Y_0,Z_0}\phi\in\Delta_3$. Then $\lr{X\cap X_0,Y\cap Y_0,(Z'\cap Y_0)\cup(Z_0\cap Y)\cup(Z'\cap Z_0)}\phi\in\Delta_2$. Recall that $Z,Z'\sub Y$, it follows that 
        $$\lr{X\cap X_0,Y\cap Y_0,((Z\cup Z')\cap Y_0)\cup(Z_0\cap Y)\cup((Z\cup Z')\cap Z_0)}\phi\in\Delta_1.$$
        Thus $\Delta_1 R^p_{\alpha}(X,Y,Z\cup Z')\Delta_3$ and (4) holds.
    \end{proof}

    \begin{definition}[$\L_\alpha$-Pre-model]\label{def:premodel}
        An $\L_\alpha$-pre-model is a set $F$ of $\L_\alpha$-MCSs such that
        for all $X,Y,Z\sub V_\alpha$ and $\Delta\in F$, the following statement holds: 
        \begin{enumerate}
            \item[(\dag)] If $\lr{X,Y,Z}\phi$ is a saturated formula in $\Delta$, then there is $\Sigma\in F$ such that $\Delta R^p_{\alpha}(X,Y,Z)\Sigma$, $\phi\in\Sigma$ and $\Sigma R^p_{\alpha}(X,Y,\ve)\Delta$.
        \end{enumerate}
        We say $\phi$ is satisfied in $F$ if there is some $\Delta\in F$ such that $\phi\in\Delta$.
    \end{definition}

    \begin{lemma}
        For each satisfiable $\phi\in\L^\preceq$, $\phi$ is satisfied in some pre-model.
    \end{lemma}
    \begin{proof}
        Let $\M=(W,\sim,\leq,V)$ be a RPD-model and $w\in W$ such that $\M,w\md\phi$. Then we define $F_\M=\{\Delta_w:w\in\M\text{ and }\Delta_w=\{\phi\in\L_\alpha:\M,w\md\phi\}\}$. It suffices to show that $F_\M$ satisfies (\dag). Suppose $\lr{X,Y,Z}\phi$ is a saturated formula in $\Delta_w$. Then $\M,w\md\lr{X,Y,Z}\phi$ and there is $u\in R(X,Y,Z)(w)$ such that $\M,u\md\phi$. Note that $\lr{X,Y,Z}\phi$ is a saturated formula, we have $w\in R(X,Y,\ve)(u)$. Then it is not hard to verify that $\Delta_w R^p_{\alpha}(X,Y,Z)\Delta_u$ and $\Delta_u R^p_{\alpha}(X,Y,\ve)\Delta_w$. Recall that $\phi\in\Delta_u$, we see that (\dag) holds for $F_\M$.
    \end{proof}

    \begin{definition}[Induced Model]
        Let $F$ be a pre-model. An $F$-path is a tuple $\tup{\Sigma_0,\psi_0,\cdots,\Sigma_{n-1},\psi_{n-1},\Sigma_n}$ where the following conditions hold for all $i<n$:
        \begin{itemize}
            \item $\psi_i=\lr{X_i,Y_i,Z_i}\phi_i$ is a saturated formula in $\Sigma_i\in F$;
            \item $\phi_i\in\Sigma_{i+1}\in F$, $\Sigma_{i+1}R^p_{\alpha}(X_i,Y_i,\ve)\Sigma_i$ and $\Sigma_i R^p_{\alpha}(X_i,Y_i,Z_i)\Sigma_{i+1}$.
        \end{itemize}
        The RPD-model $\M^F=(W^F_\Gamma,\leq^F,\sim^F,V^F)$ induced by $\Gamma\in F$ is defined by:
        \begin{itemize}
            \item $W^F_\Gamma$ is the set of all paths in $F$ begins with $\phi$.
            \item for all $y\in V_\alpha$ and $\pi,\pi'\in W^F_\Gamma$, $\pi\leq_y\pi'$ iff one of the following holds:
            \begin{itemize}
                \item $\pi'=\tup{\pi,\lr{X,Y,Z}\phi,\Sigma}$ and $y\in Y\cup Z$;
                \item $\pi=\tup{\pi',\lr{X,Y,Z}\phi,\Sigma}$ and $y\in Y$.
            \end{itemize}
            Let $\leq^F_y$ be the reflexive-transitive closure of $\leq_y$.

            \item for all $s\in V_\alpha$ and $\pi,\pi'\in W^F_\Gamma$, $\pi\rightharpoonup_s\pi'$ if and only if $\pi'=\tup{\pi,\lr{X,Y,Z}\phi,\Sigma}$ and $D_Xs\in\mathrm{last}(\pi)$. Let $\sim^F_s$ the reflexive-symmetric-transitive closure of $\rightleftharpoons_s$.
            
            \item for all $P\x\in\L_\alpha$, $V^F(P\x)=\{\pi\in W^F: P\x\in\mathrm{last}(\pi)\}$.
        \end{itemize}
    \end{definition}

    One may notice now that the construction of the desired model is almost the same as the one we used in the proof of Completeness Theorem. And similar to the proof of Completeness Theorem, with the help of Fact \ref{fact:MCSalpha} and the definition of pre-models, we can verify that the following lemma holds:

    \begin{lemma}[Truth Lemma]
        For each formula $\phi\in\L_\alpha$ and path $\pi\in W^F$, 
        \begin{center}
            $\M^F,\pi\md\phi$ if and only if $\phi\in\mathrm{last}(\pi)$.
        \end{center}
    \end{lemma}

    As a consequence, for each $\phi\in\L^\preceq$, $\phi$ is satisfiable if and only if $\phi$ is satisfied in some $\L_\phi$-pre-model. Recall that up to modal equivalence, $\L_\phi$ contains finitely many formulas, $\mathrm{MCS}_\phi$ is finite for each $\phi\in\L^\preceq$, we obtain the following theorem:

    \begin{theorem}
        The satisfiability problem of LPFD is decidable.
    \end{theorem}

\subsection{The Hybrid Extension of LPFD}\label{sec:hybridExt}

    In this subsection, we extend LPFD with nominals. By a vocabulary with nominals we mean a tuple $(\mathsf{V},\mathsf{Pred},\mathsf{Nom},\mathsf{ar})$ where $(\mathsf{V},\mathsf{Pred},\mathsf{ar})$ is a vocabulary and $\mathsf{Nom}=\set{i_k:k\in\omega}$ a denumerable set of nominals.

    The language $\mathcal{L}^\preceq_\mathsf{Nom}$ with nominals is given by:
\[
    \L^\preceq_\mathsf{Nom}\ni \phi ::= P\x \mid D_Xy \mid i \mid \neg\phi \mid \phi\wedge\phi \mid \Lr{X,Y,Z}\phi,
\]
which only differs from the language of LPFD in those nominals.

We modify the valuation $V$ in a RPD-model $\M=(W,\sim,\leq,V)$ correspondingly such that $V{\rsto}\mathsf{Nom}$ is a partial function from $\mathsf{Nom}$ to $W$. The resulted RPD-models are called RPDN-models.
The semantic truth of the nominals in an RPDN-model is defined as follows:
\begin{center}
    \begin{tabular}{lll}
        $\M,w\models i$ & if and only if & $w=V(i)$\\
    \end{tabular}
\end{center}
As usual, we call LPFD with nominals `hybrid LPFD', abbreviated to HLPFD. Let $\mathsf{Nom}$ be a fixed set of nominals. We present here the calculus $\mathsf{C}_\mathsf{Nom}$ for HLPFD and show its soundness and completeness.
Let $X,Y,Z\in\mathcal{P}^{<\aleph_0}(\mathsf{V})$, $\phi,\psi\in\L^\preceq_\mathsf{Nom}$, $i,j\in\mathsf{Nom}$, $P\in\mathsf{Pred}$ and $v\in \mathsf{V}$. The axioms and rules of $\mathsf{C}_\mathrm{HLPFD}$ are as follows:
\begin{enumerate}[(I)]
    \item[(Tau)] Axioms and rules for classical propositional logic;
    \item[(Nec)] from $\phi$ infer $\Lr{X,Y,Z}\phi$;
    \item[(K)] $\Lr{X,Y,Z}(\phi\to\psi)\to(\Lr{X,Y,Z}\phi\to\Lr{X,Y,Z}\psi)$;
    \item[(Dep)] $\phi\to\D_X\phi$, provided $\phi\in\mathsf{Atom}(X)=\set{P\x:\mathsf{set}(\x)\sub X}$;
    \item[(Nom)] $@_i\phi\to\Lr{\ve,\ve,\ve}(i\to\phi)$, provided $i\in\mathsf{Nom}$;
    %\item[(Sur)] $@_i\top$
    \item[(Name)] from $i\to\phi$ infer $\phi$, provided that $i\not\in\phi$, i.e., $i$ does not occur in $\phi$;
    \item[(Paste)] from $@_i\lr{X,Y,Z}j\to@_j\phi$ infer $@_i\Lr{X,Y,Z}\phi$, provided $i\neq j$ and $j\not\in\phi$;
    \item[] 
    \item[(DD)] Axioms and rules for $\Lr{\ }-D$ interaction:
    \item[(1)] $D_Xs\wedge\Lr{\{s\},\ve,\ve}\phi\to\Lr{X,\ve,\ve}\phi$;
    \item[(2)] $i\wedge\neg D_Xs\to\lr{X,\ve,\ve}\Lr{s,\ve,\ve}\neg i$.
    \item[] 
    \item[(Ord)] Axioms for the preference orders:
    \item[(1)] $\Lr{X,Y,\ve}\phi\to\phi$;
    \item[(2)] $\phi\to\Lr{\set{v},\ve,\ve}\lr{\set{v},\ve,\ve}\phi$;
    \item[(3)] $\lr{X,Y,Z}\lr{X',Y',Z'}\phi\to\lr{X\cap X', Y\cap Y',(Z\cap Y')\cup(Z\cap Z')\cup(Y\cap Z')}\phi$;
    \item[(4)] $@_i\lr{\ve,\ve,\{v\}}j\leftrightarrow@_i\lr{\ve,\{v\},\ve}j\wedge@_j\neg\lr{\ve,\{v\},\ve}i$, provided $i,j\in\mathsf{Nom}$;
    \item[(5)] $\lr{X,Y,Z}i\wedge\lr{X',Y',Z'}i\leftrightarrow\lr{X\cup X',Y\cup Y',Z\cup Z'}i$, provided $i\in\mathsf{Nom}$.
\end{enumerate}

%In what follows, we write $\mathsf{C}_\mathsf{Nom}$ for $\mathsf{C}_\mathrm{HLPFD}$. Soundness of $\mathsf{C}_\mathsf{Nom}$ is not hard to verify. 
Comparing $\mathsf{C}_\mathsf{Nom}$ with $\mathsf{C}$,  in addition to the standard axioms and rules for nominals, 
axioms (Ord,4,5) and (DD,2) are new, which characterize RPD-models in a more refined way. 
Note also that some old axioms in $\mathsf{C}$ are presented in $\mathsf{C}_\mathsf{Nom}$ in a different way. 
For example, axiom  (DD,1) in $\mathsf{C}_\mathsf{Nom}$ are bottom-up versions of axioms (Dep,d) in $\mathsf{C}$.

With the above mentioned changes in $\mathsf{C}_\mathsf{Nom}$ due to the addition of nominals, 
the completeness of $\mathsf{C}_\mathsf{Nom}$ can be proved by directly using the canonical model, which is a standard method and relatively routine. 
So we relegate the details of the following theorem's proof in the appendix. 
\begin{theorem}
    $\mathsf{C}_\mathsf{Nom}$ is sound and strongly complete.
\end{theorem}
Note that the equivalence between PD-models with nominals (where $I(i) \in A$ for $i\in\mathsf{Nom}$) and RPDN-models with respect to $\mathcal{L}^\preceq_\mathsf{Nom}$
can be established as in Section \ref{sec:semEqu}. 
So for the class of PD-models with nominals we also have the soundness and strong completeness of $\mathsf{C}_\mathsf{Nom}$.

As for the decidability of HLPFD, we cannot prove it by directly following the strategy used in the proof of LPFD's decidability. 
We will not attack this problem in this paper but rather leave it for future work.

\section{LPFD for Coalitional Power in Cooperative Games}\label{sec:CGinLPFD}

Coalition logic is proposed to reason about coalitional effectiveness in games in strategic form. 
However, in non-cooperative games, each player plays separately rather than as an integral part of a coalition.
The so-called coalitional effectiveness is essentially the effectiveness of an agglomeration of actions.

Should there be any difference between a coalitional action and an agglomeration of actions?
This is a key issue in the philosophical analysis of collective agency \cite{sep-shared-agency}.
In this section, we provide a game theoretical perspective on this issue by 
modelling cooperative games in strategic and coalitional form \cite[Section 11]{peleg2007Introduction} in LPFD and 
characterizing one of its solution concepts, the core, in HLPFD.

\subsection{Cooperative Games in LPFD}

Different from non-cooperative games, in cooperative games in strategic and coalitional form \cite[Section 11]{peleg2007Introduction}, 
players can not only act individually but also choose to join a coalition and act as a part of the coalition.  
In such games, the players in a coalition can do something together in agreement rather than separately. 
So coalitional actions and power are different from an agglomeration of actions and its effectiveness.
This difference is essential to our game theoretical perspective on collective agency, which we will elaborate on in Section \ref{sec:CAandPO}. 
In this part, we propose a framework based on LPFD to represent cooperative games and make the difference explicit. 

For simplicity, we restrict ourselves to the case where $\mathsf{V}$ is finite. 
We use $N = \set{1,2,\ldots,n}$ rather than $\mathsf{V}$ to indicate the finiteness of the players.

We first specify what constitutes $O$ and $A$.
To explicitly model coalitions as a different part of each player's choices from strategies, 
we distinguish between the terms ``strategy''(or equivalently ``actions'') and ``choices''.
\begin{definition}[Players' Strategies]
    Let $\Sigma$ be the set of all possible strategies of all players and 
    $\Sigma^{<N} := \bigcup_{i=1}^n \Sigma^i$ be 
    the set of all strategy sequences of length at most $n$.
\end{definition}
\begin{definition}[Players' Choices and Choices Merging]\label{def:Pchoices}
    The set of the players' choices is defined as follows:
    $$O := \set{f: I\rightarrow \Sigma\mid I\subseteq N} \enspace .$$ 
    For $f,f'\in O$ with $\dom{f}\cap \dom{f'} = \emptyset $, $f\oplus f':= f\cup f'$.
\end{definition}
For example, given three players $N = \set{1,2,3}$ and the players' possible strategies in $\Sigma = \set{\alpha,\beta}$,
$f = \set{(1,\alpha),(3,\beta)}\in O$ denotes a possible choice of the players 1 and 3 as a coalition;
$f' = \set{(2,\alpha)}\in O$ denotes a possible choice of the player 2. 
Then  $f\oplus f' =  \set{(1,\alpha),(2,\alpha),(3,\beta)}$.

In a PD-model, there is no requirement on $A\subseteq O^N$. 
This is not the case any longer when the players' choices concern forming coalitions. 
We impose three conditions on a realizable choice profile. First of all, a player cannot choose to form a coalition she is not in. 
Second, a player cannot choose to form a coalition without the others in the coalition making the same choice. 
Third, once a coalition forms, it acts as a whole, which means that its members act according to a unique strategy sequence. 
This strategy sequence can be seen as a collective plan which is made effective by common consent.

To make the definition of realizable choice profiles precise, we make use of the following notations.
\begin{notation}
    \begin{itemize}
        \item $\Pi(N)$ is the set of all partitions of $N$.
        \footnote{A partition of $N$ is a set of non-empty subsets of $N$ whose union is $N$ and which do not intersect each other.}
        \item  Given $a\in O^N$,  
        \begin{itemize}
            \item $a_i$ denotes the $i$th element of $a$, which is a function;
            \item $a_\mathsf{rng} := \set{a_i\in O\mid i\in N}$; 
            \item $a_\mathsf{dom} := \set{\dom{a_i}\subseteq N\mid i\in N}$;
            %\item $\dom{\vec{a}} := \seq{\dom{a_i},\dom{a_2},\ldots,\dom{a_n}}$.
        \end{itemize}
    \end{itemize}   
\end{notation} 

\begin{definition}[Realizable Choice and Strategy Profiles]\label{def:realizable}
    A choice profile $a\in O^N$ is realizable if and only if it satisfies the following three conditions:
    \begin{enumerate}
        \item $i\in \dom{a_i}$;
        \item $a_\mathsf{dom}\in \Pi(N)$
        \item $\dom{a_i} = \dom{a_j}$ implies that $a_i = a_j$ for all $i,j\in N$.
    \end{enumerate}
    Let $\Xi$ denote the set of all realizable choice profiles. 

        Let $a_\mathsf{merge} := \bigoplus_{f\in a_\mathsf{rng}} f$ for $a\in \Xi$. Given $A\subseteq \Xi$,  
    the set of all realizable strategy profiles of a partition $\pi\in \Pi(N)$ in $A$ is 
    $$\sigma_A(\pi) := \set{a_\mathsf{merge}\mid a\in A \text{ and } a_\mathsf{dom} = \pi}\enspace .$$
    When there is no danger of ambiguity, we will leave out the subscript $A$. 
\end{definition}

Having defined $O$ and $\Xi$, we define a class of PD-models we will work with.  

\begin{definition}[Coalition-preference-dependence (CPD) models]\label{def:CPDmodel}
    A coalition-preference-dependence model is a PD-model $\mathbb{M} = ((O,I),A)$ in which $O$ is defined in Definition \ref{def:Pchoices} and 
    $A$ and $\preceq_i$ satisfy the following conditions: 
    \begin{enumerate}
        \item $A\subseteq \Xi$;
        \item $\set{a_\mathsf{dom}\mid a\in A} = \Pi(N)$;
        \item if $\pi\in \Pi(N)$ is finer than $\pi'\in \Pi(N)$,\footnote{That is, for all $X\in \pi$ there is $X'\in \pi'$ such that $X\subseteq X'$.}
        then $\sigma_A(\pi)\subseteq \sigma_A(\pi')$;
        \item if $a_\mathsf{merge} = a'_\mathsf{merge}$, then $a\simeq_i a'$ for all $i\in N$;
        \item $\preceq_i$ is total for all $i\in N$.
    \end{enumerate}
\end{definition}
The first condition says that $A$ should contain realizable choice profiles.
The second condition says that the players can form coalitions according to all possible partitions of $N$. 
The third condition requires bigger coalitions to have no less strategies than smaller coalitions. 
The fourth condition requires that the players' preference relations depend directly on strategy profiles. 
The players' choices of coalitions can only influence the players' preferences by affecting their strategies.
The last condition requires the players' preference relations to be total, which is a standard assumption in game theory.

The following example illustrates our notations and the CPD-models. 
\begin{example}\label{exp:CPDmodel}
    Let $N = \set{1,2,3}$ and $\Sigma = \set{\alpha,\beta,\gamma}$. 
    $A$ is given in Table \ref{tab:AinCPD}. 
    \begin{table}
        \centering
        \begin{tabular}{llll}
           \, & 1                   & 2                 & 3                         \\
            $a$ & $\set{(1,\alpha)}$   & $\set{(2,\beta)}$  & $\set{(3,\alpha)}$        \\
            $a'$ &$\set{(1,\alpha),(2,\beta)}$    & $\set{(1,\alpha),(2,\beta)}$  & $\set{(3,\alpha)}$ \\
            $a''$ &$\set{(1,\alpha),(2,\gamma)}$    & $\set{(1,\alpha),(2,\gamma)}$  & $\set{(3,\beta)}$ \\
            $a^{3\prime}$ &$\set{(1,\alpha)}$    & $\set{(2,\beta),(2,\alpha)}$  & $\set{(2,\beta),(2,\alpha)}$ \\
            $a^{4\prime}$ &$\set{(1,\beta)}$    & $\set{(2,\beta),(2,\gamma)}$  & $\set{(2,\beta),(2,\gamma)}$\\
            $a^{5\prime}$ &$\set{(1,\alpha),(3,\alpha)}$    & $\set{(2,\beta)}$  & $\set{(1,\alpha),(3,\alpha)}$ \\
            $a^{6\prime}$ &$\set{(1,\gamma),(3,\alpha)}$   & $\set{(2,\alpha)}$  & $\set{(1,\gamma),(3,\alpha)}$ \\
            $a^{7\prime}$ &$\set{(1,\alpha),(2,\beta),(3,\alpha)}$    & $\set{(1,\alpha),(2,\beta),(3,\alpha)}$  & $\set{(1,\alpha),(2,\beta),(3,\alpha)}$ \\
            $a^{8\prime}$ &$\set{(1,\alpha),(2,\gamma),(3,\beta)}$    & $\set{(1,\alpha),(2,\gamma),(3,\beta)}$  &  $\set{(1,\alpha),(2,\gamma),(3,\beta)}$ \\
            $a^{9\prime}$ &$\set{(1,\beta),(2,\beta),(3,\gamma)}$    & $\set{(1,\beta),(2,\beta),(3,\gamma)}$  & $\set{(1,\beta),(2,\beta),(3,\gamma)}$ \\
            $a^{10\prime}$ &$\set{(1,\gamma),(2,\alpha),(3,\alpha)}$   & $\set{(1,\gamma),(2,\alpha),(3,\alpha)}$  & $\set{(1,\gamma),(2,\alpha),(3,\alpha)}$ \\
            $a^{11\prime}$ &$\set{(1,\gamma),(2,\gamma),(3,\gamma)}$    & $\set{(1,\gamma),(2,\gamma),(3,\gamma)}$   & $\set{(1,\gamma),(2,\gamma),(3,\gamma)}$  \\
        \end{tabular}
        \caption{$A$ in Example \ref{exp:CPDmodel}}\label{tab:AinCPD}
    \end{table}
    According to our notation, 
    \begin{itemize}
       % \item $\vec{a}'(1) = \vec{a}''(1) = \seq{\set{1,2},\set{3}}$;
        \item $a_\mathsf{merge} = a'_\mathsf{merge} = a_\mathsf{merge}^{3\prime} = a_\mathsf{merge}^{5\prime} = a_\mathsf{merge}^{7\prime} = \set{(1,\alpha),(2,\beta),(3,\alpha)}$;
        \item $\sigma(\set{\set{1},\set{2},\set{3}}) = \set{\set{(1,\alpha),(2,\beta),(3,\alpha)}}$\\
        and $\sigma(\set{\set{1,2},\set{3}}) = \set{ \set{(1,\alpha),(2,\beta),(3,\alpha)}, \set{(1,\alpha),(2,\gamma),(3,\beta)}}$. 
    \end{itemize}
    As the readers can verify, all the requirements of a CPD-model concerning $A$ are satisfied here. 
    For example, $\sigma(\set{\set{1},\set{2},\set{3}})\subseteq \sigma(\set{\set{1,2},\set{3}})\subseteq \sigma(\set{N})$.
    To make sure $\preceq_i$ satisfy the requirements, $a\simeq_i a'\simeq_i a^{3\prime}\simeq_i a^{5\prime}\simeq_i a^{7\prime}$ needs to be the case.

\end{example}

As can be easily spotted in the above example, coalitions are explicitly incorporated into the players' choices in the CPD-models. 
Once a coalition forms, the players in it act as a whole. 
Moreover, a coalition could possibly do more than its constituent parts. 

The coalition partition formed in a game directly affects each player's strategy. 
Hence it has a substantial influence on the final outcome of the game. 
Can the language of LPFD express what partition is formed in a realizable choice profile?
The following proposition gives a partially positive answer.
\begin{proposition}
    Let $\mathbb{M}=((M,A),\leq)$ be a CPD-model with $\mathbb{M},a'\models \neg D_X(-X)$ for all $a'\in A$ satisfying $a'_\mathsf{dom} = \set{X,-X}$. Then for all $a\in A$ and non-empty subset $X\sub N$, the following two are equivalent:
    \begin{enumerate}
        \item $X\in a_\mathsf{dom}$;
        \item $\mathbb{M},a\models \bigwedge_{i\in X} D_iX\wedge \bigwedge_{j\notin X} \neg D_X j$.
    \end{enumerate}
\end{proposition}
\begin{proof}
    \textbf{From 1 to 2.}

    Assume $X\in a_\mathsf{dom}$. Suppose $a'\in A$ and $a =_i a'$ for some $i\in X$. Then $X\in a'_\mathsf{dom}$. Since $A\sub\Xi$, $a_i = a_j$ and $a'_i = a'_j$ for all $i,j\in X$. Note that $a =_i a'$ for some $i\in X$, we see $a_j = a_i = a'_i = a'_j$ for all $j\in X$, i.e. $a =_X a'$. Thus $\mathbb{M},a\models D_iX$. By the arbitrariness of $i\in X$, we see $\mathbb{M},a\models \bigwedge_{i\in X} D_iX$.
    
    When $X=N$, we see that $\bigwedge_{j\notin X} \neg D_X j$ is $\top$ and $\mathbb{M},a\models\bigwedge_{j\notin X} \neg D_X j$. Suppose $X\neq N$. Take an arbitrary $j\not\in X$.
    Then we have the following cases:
    \begin{itemize}
        \item $a_\mathsf{dom}\neq\set{X,-X}$. Let $\pi = \set{X,-X}$. Note that $\sigma_A(a_\mathsf{dom})\subseteq \sigma_A(\pi)$, there must be $b\in A$ such that $b_\mathsf{dom} = \pi$ and $a_\mathsf{merge} = b_\mathsf{merge}$. Then it must be the case that $\dom{b_j}= -X\neq \dom{a_j}$ and so $a \neq_j b$.
        \item $a_\mathsf{dom}=\set{X,-X}$. Since $\mathbb{M},a\models \neg D_X(-X)$, there must be $b\in A$ such that $a =_X b$ and  $a\neq_{-X} b$. If $a_\mathsf{dom} \neq b_\mathsf{dom}$, then $\dom{a_j}= -X\neq \dom{b_j}$ and so $a\neq_j a'$. Suppose $a_\mathsf{dom} = b_\mathsf{dom}$. Then $a_k$ are all the same for $k\in -X$ and $a'_h$ are all the same for $h\in -X$. Since $a\neq_{-X} a'$, we see $a_j \neq a'_j$.
    \end{itemize}
    Hence $\mathbb{M},a\models \neg D_X j$.
    By the arbitrariness of $j$, we see $\mathbb{M},a\models \bigwedge_{j\notin X} \neg D_X j$.

    \textbf{From 2 to 1.}

    Assume that $X\not\in a_\mathsf{dom}$ and $\mathbb{M},a\models \bigwedge_{i\in X} D_iX\wedge \bigwedge_{j\notin X} \neg D_X j$. Let $x\in X$.
    \begin{itemize}
        \item $X\subsetneq\mathsf{dom}(a_x)$. Then there is $j\in \mathsf{dom}(a_x)\setminus X$ such that $a_j = a_i$ for all $i\in \mathsf{dom}(a_x)$. So for all $a' =_X a$, $a'_j = a_i = a'_i$ for all $i\in X$. Then we have $\mathbb{M},a\models  D_X j$ where $j\notin X$. Contradiction!
        \item Otherwise, there is $j\in X\setminus \mathsf{dom}(a_x)$. Since $\mathbb{M},a\md D_xX$, we see $\mathbb{M},a\md D_{\mathsf{dom}(a_x)}j$. By the direction we have proved above, $\mathbb{M},a\md\neg D_{\mathsf{dom}(a_x)}j$, which is a contradiction.
    \end{itemize}
\end{proof} 

The assumption of the above proposition that $\mathbb{M},a\models \neg D_X(-X)$ for all $a\in A$ satisfying $a'_\mathsf{dom} = \set{X,-X}$ requires that 
no coalition can completely decides what its complementary coalition chooses to do. 
If $X$ can completely control what $-X$ chooses, 
then the division of $X$ and $-X$ is senseless, because $\mathbb{M},a\models  D_X N$ follows from $\mathbb{M},a\models D_X -X$.
As the readers can verify, the CPD-model in Example \ref{exp:CPDmodel} does not satisfy the assumption at $a'',a^{4\prime},a^{6\prime}$. 

To avoid vacuous coalitions division, we will work with the CPD-models with the above assumption. 
\begin{definition}[Real CPD-models]\label{def:RCPDmodel}
    A real CPD-model (RCPD-model) $\mathbb{M}$ is a CPD-model that satisfies the assumption that 
    $\mathbb{M},a\models \neg D_X(-X)$ for all $a\in A$ satisfying $a_\mathsf{dom} = \set{X,-X}$.
\end{definition}
In a RCPD-model, $\bigwedge_{i\in X} D_iX\wedge \bigwedge_{j\notin X} \neg D_X j$ expresses that $X$ is in the coalition partition. 
We will use the abbreviation 
$$p_X := \bigwedge_{i\in X} D_iX\wedge \bigwedge_{j\notin X} \neg D_X j$$ 
for convenience in the next section, 
where we demonstrate that LPFD as presented in this section provides a useful scaffolding for approaching several issues on coalitions.

\subsection{The Core in HLPFD}

Having set up the LPFD framework for representing cooperative games in strategic and coalitional games, 
in this part, we show that the core, an important solutions concept in the cooperative game theory, can be expressed in HLPFD.
Moreover, by considering functional dependence explicitly, 
we generalize the core and show how it is related to Nash equilibrium and Pareto optimality.

Just as Nash equilibrium in non-cooperative games captures stability of a strategy profile,
the concept of the core, as a basic solution concept in cooperative games, also captures stability of a strategy profile in cooperative games. 
The difference is that the core takes the stability of a coalition into consideration. 
There are other notions for characterizing stability in cooperative games, for example, stable set, bargaining set and so on. 
In this paper, we focus on the core.

The concept of the core is formulated in CPD-models as follows. \footnote{
    The definition of the core can vary in different settings. 
    Our definition is based on \cite[Definition 2.2]{conzalez2021coreaxiom}, which is a relatively general version.
    }
\begin{definition}[Core in CPD-Model] \label{def:theCore}
    Given a CPD-model $\mathbb{M}$, a choice profile $a\in A$ is in the core of $\mathbb{M}$ if and only if 
    \begin{enumerate}
        \item $a_\mathsf{dom} = \set{N}$; and 
        \item there is no $X\subseteq N$ and $a'\in A$ such that 
            \begin{enumerate}
                \item $X\in a'_\mathsf{dom}$; and
                \item for all $a'' =_X a'$ and all $i\in X$, $a \prec_i a''$.
            \end{enumerate}
    \end{enumerate}
    Let $Co_\mathbb{M}$ denote the core of $\mathbb{M}$.
\end{definition}
If $N$ arrives at a choice profile $a$, which is in the core, 
then no $X\subset N$ has any incentive to deviate from the coalition $N$, 
because forming the coalition $X$ cannot guarantee all players in $X$ end up with a better outcome. 
Coalitional power plays a key role in the basic idea of core, 
because whether $X$ has any incentive to deviate depends on
whether $X$ as a coalition can force a choice profile that all of its members prefer to the current choice profile.

Note that according to the definition of the core, if $X = N$, there is no other choice profile with the coalition partition $\set{N}$ 
which is strictly preferred by every player in $N$. 
Namely, $a$ is weakly Pareto optimal among the choice profiles with the coalition partition $\set{N}$. 
In fact, the following proposition holds.
\begin{proposition}\label{prop:coreToPareto}
    Given a CPD-model $\mathbb{M}$, if a choice profile $a\in A$ is in the core of $\mathbb{M}$ then $a$ is weakly Pareto optimal. 
\end{proposition}
\begin{proof}

    %For all $a\in A$ with $a_\mathsf{dom} = \set{N}$, $a_i = a_j$ for all $i\neq j\in N$ and 
    %any player cannot change its choice without causing the other players to change their choice. 
    %So the definition of Nash equilibrium is trivially satisfied by such a choice profile.

    Since $\mathbb{M}$ satisfies the condition that $\sigma_A(\pi)\subseteq \sigma_A(\set{N})$ for all $\pi\in \Pi(N)$,
    by the fourth condition of Definition \ref{def:CPDmodel},
    the weak Pareto optimality of $a$ within the choice profiles having $\set{N}$ as their coalition partition can be generalized trivially to all choice profiles.  
\end{proof}

The following example illustrates the concept of core and how it differs from Nash equilibrium and Pareto optimality. 
\begin{example}\label{exp:prisonercore}
    Let $N = \set{1,2}$ and $\Sigma = \set{\alpha,\beta}$. 
    $A$ and the preference relations  are given in Table \ref{tab:prisonercore}. 
    The preference relations are given in the form of a pair of ordinal utilities where the first element is for player 1 and the second for player 2.
    \begin{table}
        \centering
        \begin{tabular}{llll}
           \, & 1                   & 2                   & Ordinal Utility                   \\
            $a$ & $\set{(1,\alpha)}$   & $\set{(2,\alpha)}$     & (9,9)   \\
            $a'$ &$\set{(1,\alpha)}$ & $\set{(2,\beta)}$    & (0,10)\\
            $a''$ &$\set{(1,\beta)}$ & $\set{(2,\alpha)}$  & (10,0)  \\
            $a^{3\prime}$ &$\set{(1,\beta)}$ & $\set{(2,\beta)}$  & (1,1)  \\  
            $a^{4\prime}$ &$\set{(1,\alpha),(2,\alpha)}$   & $\set{(1,\alpha),(2,\alpha)}$ &  (9,9) \\
            $a^{5\prime}$ &$\set{(1,\alpha),(2,\beta)}$     & $\set{(1,\alpha),(2,\beta)}$ & (0,10) \\
            $a^{6\prime}$ &$\set{(1,\beta),(2,\alpha)}$     & $\set{(1,\beta),(2,\alpha)}$ & (10,0) \\
            $a^{7\prime}$ &$\set{(1,\beta),(2,\beta)}$     & $\set{(1,\beta),(2,\beta)}$  & (1,1)\\
        \end{tabular}
        \caption{$A$ in Example \ref{exp:prisonercore}}\label{tab:prisonercore}
    \end{table}
   Readers familiar with game theory can recognize that without the last four rows the table represents the prisoners' dilemma. 
   $a^{3\prime}$ is a Nash equilibrium but $a$ is not as in the original prisoners' dilemma. 
   Now our coalitional version allows player 1 and player 2 to form a coalition by whatever means, 
   for example, a binding agreement or switching to the mode of team reasoning simultaneously. 
   So there are four extra profiles in which both players explicitly choose to join the coalition.  
   Among these four extra profiles, $a^{4\prime}$ is the only element in the core. 
   Thus it is both Pareto optimal and a Nash equilibrium. 

   Note that in the example $\set{1,2}$ as a coalition does not expand what each of the players can choose, 
   namely $\sigma(\set{1,2}) = \sigma(\set{\set{1},\set{2}})$.
   But it still makes some difference. 
   This difference is brought about by something collective as clearly reflected in our example.
   The core captures this collective element in the example.
\end{example}

Next, we show that the core can be expressed in HLPFD with respect to the class of RCPD-models (Definition \ref{def:RCPDmodel}) with nominals. 
\begin{proposition}\label{prop:theCoreinRCPDN}
    Given a RCPD-model $\mathbb{M}$ with nominals $\mathsf{Nom}$,  the current choice profile $a$ with name $i$, i.e., $a = I(i)\in A$, is in the core of $\mathbb{M}$,
    if and only if 
        $$\mathbb{M},a\models i\wedge p_N\wedge \bigwedge_{\emptyset\neq X\subseteq N} \invertedforall (p_X\rightarrow \lr{X,\emptyset,\emptyset}\bigvee_{x\in X}\lr{\emptyset,\{x\},\emptyset}i).$$
\end{proposition}

In fact, as in the case of Nash equilibrium and Pareto optimality, we can also have a relativized version of the core as follows
$$\mathsf{Core}_X i :=  i\wedge p_X\wedge \bigwedge_{\emptyset\neq C\subseteq X} \mathbb{D}_{-X} (p_C\rightarrow \lr{-X\cup C,\emptyset,\emptyset}\bigvee_{c\in C}\lr{-X,\set{c},\emptyset}i)$$
Note that when taking $X = N$, we get the original definition of the core as expressed in Proposition \ref{prop:theCoreinRCPDN}.
The relativized version of the core enables us to express some interesting relationships between coalitions. 
For example,
$$\mathsf{Core}_X i\wedge \mathsf{Core}_{-X} i$$
which says that in the current choice profile $i$, both $X$ and $-X$ form coalitions and are in their relativized cores.

More generally, we can define the following concept:
$$\mathsf{Core}_\pi i := \bigwedge_{X\in \pi} \mathsf{Core}_X i$$
where $\pi$ is a partition of $N$. 
It characterizes the stability of a collection of coalitions at a choice profile $i$. 
The core is a special case of it where $\pi = \set{N}$.
Moreover, Nash equilibrium $\Na N$ is also a special case of it where $a_\mathsf{dom} =\pi = \set{\set{1},\set{2},\ldots,\set{n}}$.
\begin{theorem}
    Given a RCPD-model $\mathbb{M}$ with nominals $i\in \mathsf{Nom}$, and $a\in A$ with $a_\mathsf{dom} =\pi = \set{\set{1},\set{2},\ldots,\set{n}}$,
    $$\mathbb{M},a\models \mathsf{Core}_\pi i \leftrightarrow (i\wedge \Na N)\enspace .$$
\end{theorem}
As a corollary to this proposition, we see that unlike the core $\mathsf{Core}_\pi i$ does not necessarily imply the weak Pareto optimality of $i$ for $N$.
But the following generalization of Proposition \ref{prop:coreToPareto} holds.
\begin{theorem}\label{prop:GeneralcoreToPareto}
    Given a RCPD-model $\mathbb{M}$ with nominals $i\in \mathsf{Nom}$ and $a\in A$, 
    $$\mathbb{M},a\models \mathsf{Core}_\pi i \rightarrow \bigwedge_{X\in \pi}\wPa X\enspace .$$
\end{theorem}

Therefore, in the sense of the above two theorems, 
our generalization of the core can be seen as a notion that unifies the core, 
Nash equilibrium and Pareto optimality. 

\section{Stability, Coalitional Power and Collective Agency}\label{sec:CAandPO}

In this section, we show how the CPD-models can help clarify issues on collective agency and 
explore some philosophical implications from the game-theoretical perspective on collective agency in CPD-models.

Philosophical discussions about collective agency have flourished in recent decades. 
Despite disagreements on the detailed definition of collective agency, 
most theories share the idea that joint actions by a group with collective agency are more than simply a coordination or cooperation between its members 
(cf. \cite{bratman2014}; \cite{gilbert2006theory}; \cite{list2011group}; \cite{searle2010making}; \cite{tollefsen2002organizations}; \cite{tuomela2013social}). 
Nevertheless, the conundrum is where does this essential difference lie. 
Gilbert \cite{gilbert2006theory}, Searle \cite{searle2010making}, and Tuomela \cite{tuomela2013social} admit an irreducible concept of a collective in a methodological sense. 
In Gilbert, it is a unique type of commitment of will: ``joint commitment''; 
in Searle, it is a special kind of intention: ``we-intention"; 
and in Tuomela, it is a complex of ``we-intention" and a particular form of attitude: ``we-mode." 
They all try to start from an irreducible concept of a collective to capture the extras of joint actions. 
In a similar sense, List and Pettit \cite{list2011group} emphasize that 
a group with agency must have a procedure to ensure that its decision-making process meets the necessary functional conditions of an agent, 
such as manifest rationality; 
Tuollefsen \cite{tollefsen2002organizations} highlights that a group with agency must contain stable structures in order to conform to the basic phenomena that can be reasonably explained by the observer. 
Even for Bratman \cite{bratman2014}, who famously argues that we can explain collective agency without any irreducible concept of a collective, 
he still claims the critical role of the interdependent relations and the mesh of individual plans between members in forming collective intentions.

As Bratman emphasized, we also pay special attention to the critical role of interdependence in forming a collective agent. 
Instead of intentionality, following the standard game-theoretical approach, 
individual preferences and choices are the starting point of our analysis, 
from which we explore the stability of the interdependence as embodied in the core. 

%As Bratman emphasized, we also pay particular attention to the critical role of interdependence in forming a collective agent. 
%This can be seen from the fact that in our basic dependence framework, by excluding the axiom of supperadditivity, 
%i.e., players' choices are independent of each other, 
%we insist that a member's choice in a group with agency always depends on other members.
%Along this way, our exploration will be from a game-theoretical perspective. 
%Following basic abstraction and restriction in game theory, we focus on the non-psychological aspects that make joint action collective. 
%We only presuppose individual preferences without assuming any initial concept of a collective or its intention. 
%This approach abstracts individual intentionality, but it also draws boundaries for our formal analysis. 
%On the one hand, philosophical or psychological discussions are helpful only in questioning the sources of individual preferences and applying formal analysis to specific situations. %;discussions involving spiritual aspects are outside formal analysis.
%On the other hand, by dividing this borderline, our analysis covers only facts, i.e., individual preferences and inter dependency between them as the base facts and the collective decisions derived from them. In this way, the mystical and spiritual elements in the collective context are removed.%, thus ensuring the objectivity of formal analysis.

In CPD-models, coalitions are taken explicitly as a part of each individual player's choices. 
That is, each player chooses which coalitions to join. 
This makes it possible to distinguish between a group action and a set of individual actions. 
In Example \ref{exp:prisonercore}, although $a^{4\prime}$ and $a$ have the same strategy profile, 
namely $a^{4\prime}_\mathsf{merge} = a_\mathsf{merge}$, 
acting together ($a^{4\prime}_\mathsf{dom} = \set{N}$) or acting individually ($a_\mathsf{dom} = \set{\set{1},\set{2}}$) make two totally different choice profiles. 
However, condition 4 in Definition \ref{def:CPDmodel} stipulates that once their strategy profile keeps the same, 
no players would prefer one to the other. 
This means that $a^{4\prime}$ and $a$ make no difference to both players' preference relations. 
Then what can the difference between $a^{4\prime}$ and $a$ bring about to the players?
The critical observation is that $a^{4\prime}$ is in the core while $a$ is not even a Nash equilibrium. 
That is, although their strategy profiles are the same, one as a joint action of the group is stable (in the sense of the core) 
while the other as cooperation of two parties is not stable (in the sense of the Nash equilibrium). 
This suggests that the stability of acting together is essential for understanding collective agency.

To elaborate on the above claim, we first make the following clarification about condition 3 in Definition \ref{def:CPDmodel}. 
It does \emph{not} require that by choosing to join the same coalition together, 
the players in the coalition should have \emph{more} strategies than they have when acting separately, 
but only \emph{no less than}. 
We leave it open whether it is a strict inclusion ($\subset$) or an equation ($=$). 
As we can see in Example \ref{exp:prisonercore}, the coalition $\set{1,2}$ does not have extra strategies. 
The special status of $a^{4\prime}$ does not rely on having a strategy profile that cannot be realized by the players separately. 
%That is to say, we do not need to presuppose redundant concepts to describe collective agency. Coalitions are those stable and undeviated states of choice, which are part of all the possible states of choice.
  
The comparison between $a^{4\prime}$ and $a$ highlights the role of the coalition in making the cooperation stable. 
It reveals that collective agency should be a binding power that makes a coalition and its joint action stable. 
This binding power may come from different sources and be present in different forms over which various theories on collective agency debate. 
No matter which source it comes from and which form it takes, the binding power should come with the stability of what it binds together. 
Regarding stability, we share the same spirit with \cite{gold2007collective}; \cite{sugden2003the}; \cite{tollefsen2002organizations}; \cite{tuomela2013social}, 
in which they also directly or indirectly take stability as a condition for the formation of a collective agent. 
Moreover, suppose we further abstractly understand the concept of the core as a specific pattern for inter-sub-coalition relations within a coalition. 
In that case, our interpretation highlights the understanding of collective agency as a relatively stable state of relations rather than an imagined conceptual entity. 
In this sense, we are in line with the call for a relationalist account (cf. \cite{baier1997doing}; \cite{meijers2003can}; \cite{schmid2003can}; \cite{wangstokhof2022}).

Game theory, especially the cooperative game theory, is a powerful tool for analyzing the kind of stability we consider essential for collective agency. 
The concept of the core is not the only solution concept in cooperative game theory. 
A lot of other solution concepts have been proposed, taking different issues related to coalitional stability into consideration.
Abstracting and logically fusing these concepts into a unified framework will bring more insights into the philosophical discussion of collective agency.
Our analysis by CPD-models serves as a first attempt to make this connection explicit by testing collective agency in games.

\section{Related Works and Conclusion}

\subsection{Related Works}

Before conclusion, we compare our work with two closely related works, 
the modal coalitional game logic (MCGL) in \cite{agotnes2009}\footnote{
    There are two logics in \cite{agotnes2009}. MCGL is the second one. The first one is more customized and limited than the second one. 
    For example, it only considers finite games where both players and states need to be finite.
}
and the logic of ceteris paribus preference (LCP) in \cite{vanBenthem2007}.

All the three works involve the modal way of modeling preference, that is, using modal operators for characterizing preorders. 
Of the three works, as regards to basic modal operators for preference, LCP is the simplest one. 
Given a preorder $\preceq$ in its semantic model, it only includes one modal operator for $\preceq$ and one for $\prec$.
MCGL concerns a multi-agent setting where for each agent there is a preorder. 
Besides modal operators for individual agents, MCGL includes group operators, 
one for the intersections of a set of preorders and one for the intersection of a set of strict preorders.
It also includes modal operators for the inverse of the preorders and a difference operator. 
Nevertheless, it does not have any operator for the intersection of strict and non-strict preorders. 
Our logic has such operators and we show that they are critical for expressing strong Pareto optimality. 

Next, with each of these two other logics, the comparison will focus on different aspects. 

\medskip

\noindent \textbf{Comparison with \cite{agotnes2009} on different formulations of the core}\\
It is shown in \cite{agotnes2009} that MCGL can express not only the core in coalitional games but also the stable set and the bargaining set. 
However, the setting they adopt for representing coalitional games is not general enough to model
the coalitional games formalized by the CPD-models. 
The limitation is due to their way of defining the coalitional effective function or the characteristic function as they call it.
In a CPD-model $\mathbb{M}$, their characteristic function can be understood as $V: 2^N\setminus \set{\emptyset} \rightarrow \mathcal{P}(A)$,  
a function assigning a set of choice profiles to each coalition.
Their formulation of the core only requires that the current choice profiles are strictly preferred to all the choice profiles in $V(X)$ for all $X\subseteq N$.
But in our formulation of the core in Definition \ref{def:theCore}, what matters is the following set for each $X\subseteq N$
$$E(X) := \set{a(X)\subseteq A\mid a\in A \text{ and } X\in a_\mathsf{dom}}$$
where $a(X) := \set{a'\in A\mid a =_X a'}$. 
$E: 2^N\rightarrow \mathcal{P}(\mathcal{P}(A))$ is a function assigning to each coalition a set of sets of choices profiles. 
This is in line with the coalitional effective function defined in subsection \ref{sec:coalitionlogic} with only one difference, 
namely $E(X)$ here is not upward closed.
Our formulation of the core requires a comparison between the current choice profile and each of the set in $E(X)$.
%As to the comparison between the current choice profile and a certain set in $E(X)$, there is no difference between our formulation and the formulation in \cite{agotnes2009}.
Note that the compartmentalization of what a coalition $X$ can enforce as $E(X)$ formalizes it is essential for our formulation of the core, 
because what a coalition $X$ can enforce depends on what $X$ would do.
This subtlety is not captured by the characteristic function in \cite{agotnes2009}.

\medskip

\noindent \textbf{Comparison with \cite{vanBenthem2007} on different ways of characterizing dependence}
We have seen that in LPFD variables are taken to partition the space of possible assignments according to their possible values.
The dependence relation is the relation between different partitions.
In LCP, what partitions the space of possible states are all possible sets of formulas of its base language.
If we think of a formula as a binary variable with its values $0$ or $1$, then the operators $[\Gamma]$, $[\Gamma]^{\preceq_x}$ and $[\Gamma]^{\prec_x}$ in LCP
correspond to our operators $\br{\Gamma,\emptyset,\emptyset}$,  $\br{\Gamma,x,\emptyset}$ and $\br{\Gamma,\emptyset,x}$ respectively.
This raises an interesting question: if we only allow binary variables, what is the difference 
between using variables (as in LFD) and formulas (as in LCP) to capture the functional dependence between variables?
Furthermore, do we really lose anything in LFD if we only allow binary variables?
A systematic study of these two questions would require future work.

\subsection{Conclusion and More Future Work}\label{sec:CandFW}

We have proposed two logics by extending LFD and studying their axiomatizations and other properties. 
We have also demonstrated how our logics can help reason about the notions of dependence, preference and coalitional power in a game theoretical setting and 
provide a unified view on three key concepts in game theory, i.e., Nash equilibrium, Pareto optimality and the core.
On the basis of the two logics, we bring novel insights to the general discussion on collective agency, where we consider agency of a collective as a stable state that is constituted by each member's preference and the interdependency between them. 
%Such a state is stabilized by the binding power of a collective, where each member is stable in her current choice and has no incentive to deviate from their joint action. 
%Through the provided unified perspective, we can clearly understand the connotation of this binding power, and we claim that it is minimally an essential feature for collective agency.

More work on collective agency from a cooperative-game-theoretical perspective needs to be done as we have instigated. 
The connection between LFD and the coalition logic we have revealed indicates that it may be fruitful to explore the relationship between LPFD and ATL \cite{goranko2004ATL}.
Some work has been done on exploring the temporal dimension of dependence \cite{dazhu2022}. 
Further work in these directions could make a logical analysis of extensive games more full-fledged.

\section*{Appendix}

\subsection*{Strong Completeness of $\mathsf{C}_\mathsf{Nom}$}

\begin{lemma}\label{lem:Lindenbaum}
    Let $\Gamma$ be a $\mathsf{C}_\mathsf{Nom}$-consistent set and $\mathsf{Nom'}=\mathsf{Nom}\cup\{j_n:n\in\omega\}$. Then $\Gamma$ can be extended to a maximal $\mathsf{C}_\mathsf{Nom'}$-consistent set $\Gamma^+$ of formulas satisfying the following conditions:
    \begin{enumerate}[-10000em]
        \item[(Named)] $\Gamma^+\cap\mathsf{Nom'}\neq\ve$;
        \item[(Pasted)] For all $@_i\lr{X,Y,Z}\phi\in\Gamma$, there is a nominal $j\in\mathsf{Nom'}$ such that $@_i\lr{X,Y,Z}j\wedge@_j\phi\in\Gamma$.
    \end{enumerate}
\end{lemma}
The proof of Lemma \ref{lem:Lindenbaum} is standard.

\begin{fact}\label{fact:nominal-model}
    Let $\Gamma$ be a named and pasted maximal $\mathsf{C}_{\mathsf{Nom}}$-consistent set. For each $i\in\mathsf{Nom}$ such that $@_i\top\in\Gamma$, let $\Delta_i=\{\phi:@_i\phi\in\Gamma\}$. Then for all $i,j\in\mathsf{Nom}$,
    \begin{enumerate}[(1)]
        \item $\Delta_i$ is a maximal $\mathsf{C}_\mathsf{Nom}$-consistent set.
        \item $i\in\Delta_j$ if and only if $\Delta_i=\Delta_j$.
    \end{enumerate}
\end{fact}

\begin{definition}\label{def:Nom-model}
    Given a named and pasted maximal $\mathsf{C}_\mathsf{Nom}$-consistent set $\Gamma$, we define the canonical model 
    %$\M_\Gamma=(W_\Gamma,R_\Gamma,V_\Gamma)$ for $\Gamma$ as follows:
    $\M_\Gamma=(W_\Gamma,\sim_\Gamma,\leq_\Gamma,V_\Gamma)$ for $\Gamma$ as follows:
    \begin{itemize}
        \item $W_\Gamma=\{\Delta_i:@_i\top\in\Gamma\text{ and }\Delta_i=\{\phi:@_i\phi\in\Gamma\}\}$;
        %\item $\Delta_iR_\Gamma(X,Y,Z)\Delta_j$ if and only if $@_i\lr{X,Y,Z}j\in\Gamma$;
        \item for each $v\in\mathsf{V}$, $\Delta_i\sim_v\Delta_j$ if and only if $@_i\lr{\set{v},\ve,\ve}j\in\Gamma$;
        \item for each $v\in\mathsf{V}$, $\Delta_i\leq_v\Delta_j$ if and only if $@_i\lr{\ve,\set{v},\ve}j\in\Gamma$;
        \item $V(P\x)=\{\Delta_i:@_iP\x\in\Gamma\}$ and $V(i)=\Delta_i$.
    \end{itemize}
\end{definition}

\begin{lemma}\label{lem:Nom-model}
    Let $\Gamma$ be a named and pasted maximal $\mathsf{C}_\mathsf{Nom}$-consistent set. Then $\M_\Gamma=(W,\sim,\leq,V)$ is an RDPN-model.
\end{lemma}
\begin{proof}
    Let $v\in\mathsf{V}$. By axiom (Ord,1,2,3), $\sim_v$ is a pre-order and $\leq_v$ is an equivalence relation. Then $(W,\sim,\leq)$ is an RPD-frame. Note that $V(i)\in W$ for each $i\in\mathsf{Nom}\cap\mathrm{dom}(V)$. To show that $\M_\Gamma$ is a RPDN-model, it suffices to show that $V$ satisfies (Val). Let $\x=(x_1,\cdots,x_n)$. Suppose $\Delta_i\sim_{\mathsf{set}(\x)}\Delta_j$ and $\Delta_i\in V(P\x)$. Then $P\x\in\Delta_i$. By (Dep), $\D_XP\x\in\Delta_i$, which entails $P\x\in\Delta_j$.
\end{proof}

\begin{lemma}\label{lem:Nom-model-properties}
    Let $\Gamma$ be a named and pasted maximal $\mathsf{C}_\mathsf{Nom}$-consistent set, $\M_\Gamma=(W,\sim,\leq,V)$, $i\in\mathsf{Nom}$ and $\Delta_i\in W$. Then
    \begin{enumerate}[(1)]
        \item If $\lr{X,Y,Z}j\in\Delta_i$, then $\Delta_iR(X,Y,Z)\Delta_j$;
        \item If $@_i\lr{X,Y,Z}\phi\in\Gamma$, then there is $j\in\mathsf{Nom}$ with $\phi\in\Delta_j$ and $\Delta_iR(X,Y,Z)\Delta_j$.
        \item $D_Xs\in\Delta_i$ if and only if $\M_\Gamma,\Delta_i\md D_Xs$.
        \item For all $\phi\in\L_\mathsf{Nom}$, $\phi\in\Delta_i$ if and only if $\M_\Gamma,\Delta_i\md\phi$.
    \end{enumerate}
\end{lemma}
\begin{proof}
    For (1), suppose $\lr{X,Y,Z}j\in\Delta_i$. By axiom (Ord,5), we see $\lr{\set{x},\ve,\ve}j,$ $\lr{\ve,\set{y},\ve}j,$ $\lr{\ve,\ve,\set{z}}j\in\Delta_i$ for all $x\in X$, $y\in Y$ and $z\in Z$, which entails by axiom (Ord,4) that $\Delta_i\sim_X\Delta_j$, $\Delta_i\leq_Y\Delta_j$ and $\Delta_i<_Z\Delta_j$. Thus $\Delta_iR(X,Y,Z)\Delta_j$.

    For (2), suppose $@_i\lr{X,Y,Z}\phi\in\Gamma$. Since $\Gamma$ is pasted, there is $j\in\mathsf{Nom}$ such that $@_i\lr{X,Y,Z}j\wedge@_j\phi\in\Gamma$. Thus $\phi\in\Delta_j$ and $\Delta_iR(X,Y,Z)\Delta_j$.

    For (3), suppose $D_Xs\in\Delta_i$ and $\Delta_i\sim_X\Delta_j$. We show that $\lr{\set{s},\ve,\ve}j\in\Delta_i$. Assume $\lr{\set{s},\ve,\ve}j\not\in\Delta_i$. Then by axiom (DD,1), we see $\D_X\neg j\in\Delta_i$, which contradicts to $\Delta_i\sim_X\Delta_j$. Thus $\M_\Gamma,\Delta_i\md D_Xs$. Suppose $D_Xs\not\in\Delta_i$. Then $i\wedge\neg D_Xs\in\Delta_i$. By axiom (DD,2), we see $@_i\lr{X,\ve,\ve}\D_s\neg i\in\Gamma$. Since $\Gamma$ is pasted, there is $j\in\mathsf{Nom}$ such that $@_i\lr{X,\ve,\ve}j\wedge@_j\D_s\neg i\in\Gamma$. Thus $\Delta_i\sim_X\Delta_j$ and $\Delta_i\not\sim_s\Delta_j$. Note that $\sim_s$ is symmetric, $\Delta_j\not\sim_s\Delta_i$. Thus $\M_\Gamma,\Delta_i\not\md D_Xs$.

    For (4), the proof proceeds by induction on the complexity of $\phi$. The case when $\phi=D_Xs$ follows from (3) immediately. The case $\phi=P\x$ or $\phi\in\mathsf{Nom}$ is trivial. The Boolean cases are also trivial. 
    Let $\phi=\Lr{X,Y,Z}\psi$. Assume $\Lr{X,Y,Z}\psi\not\in\Delta_i$. Then $\lr{X,Y,Z}\neg\psi\in\Delta_i$ and so $@_i\lr{X,Y,Z}\neg\psi\in\Gamma$. By (2), $\neg\psi\in\Delta_j$ for some $\Delta_j\in R(X,Y,Z)(\Delta_i)$. Then $\psi\not\in\Delta_j$ and by induction hypothesis, $\M_\Gamma,\Delta_j\not\md\psi$, which entails $\M_\Gamma,\Delta_i\not\md\Lr{X,Y,Z}\psi$. Assume that $\M_\Gamma,\Delta_i\not\md\Lr{X,Y,Z}\psi$. Then there is $\Delta_j\in R(X,Y,Z)(\Delta_i)$ such that $\M_\Gamma,\Delta_j\not\md\psi$. By induction hypothesis, $\psi\not\in\Delta_j$ and so $\neg\psi\wedge j\in\Delta_j$. Note that $\lr{X,Y,Z}j\in\Delta_i$, we see $\lr{X,Y,Z}\neg\psi\in\Delta_i$, which entails $\Lr{X,Y,Z}\psi\not\in\Delta_i$.
\end{proof}

% \begin{theorem}
%     $\mathsf{C}_\mathsf{Nom}$ is sound and strongly complete w.r.t RPDN-models.
% \end{theorem}
%Now we are ready to prove the soundness and completeness of $\mathsf{C}_\mathsf{Nom}$:\\

\noindent\textbf{Theorem. }{\em $\mathsf{C}_\mathsf{Nom}$ is sound and strongly complete.}
\begin{proof}
    Soundness is not hard to verify. Let $\Gamma^-\sub\L_\mathsf{Nom}$ be any consistent set of formulas. By Lemma \ref{lem:Lindenbaum}, $\Gamma^-$ can be extended to a named and pasted maximal $\mathbf{LPFD_{Nom'}}$-consistent set $\Gamma$. By Lemma \ref{lem:Nom-model}, the triple $\M_\Gamma=(W_\Gamma,R_\Gamma,V_\Gamma)$ defined in Definition \ref{def:Nom-model} is a DP-model with nominals. By Lemma \ref{lem:Nom-model-properties}, we see $\M_\Gamma\md\Gamma^-$. Then $\M_\Gamma{\rsto}\mathsf{Nom}$ is a RPDN-model satisfying $\Gamma^-$.
\end{proof}

\bibliographystyle{splncs04}
\bibliography{RDependencePreferenceCoalitionalPower_0801}

\end{document}